\documentclass[
 preprint,
 amsmath,amssymb,
 aps,
]{revtex4-1}
\usepackage{graphicx}
\usepackage{subfigure}
\usepackage{dcolumn}
\usepackage{bm}
\usepackage{color}
\usepackage{soul}

\begin{document}

\preprint{APS/123-QED}

\newtheorem{lemma}{Lemma}
\newtheorem{corollary}{Corollary}

\title{Mechanisms of a Convolutional Neural Network System for Learning Three-dimensional Unsteady Wake Flow}
\author{Sangseung Lee}
\author{Donghyun You}%
 \email{dhyou@postech.ac.kr}
\affiliation{
Pohang University of Science and Technology, 77 Cheongam-ro, Nam-gu, Pohang, Gyeongbuk 37673, Republic of Korea
}%
\date{\today}
\begin{abstract}
In the present study, a convolutional neural network (CNN) system which consists of multiple multi-resolution CNNs to predict future three-dimensional unsteady wake flow using flow fields in the past occasions is developed.
Mechanisms of the developed CNN system for prediction of wake flow behind a circular cylinder are investigated in two flow regimes: the three-dimensional wake transition regime and the shear-layer transition regime.
Understanding of mechanisms of CNNs for learning fluid dynamics is highly necessary to design the network system or to reduce trial-and-errors during the network optimization.   
Feature maps in the CNN system are visualized to compare flow structures which are extracted by the CNN system from flow in the two flow regimes. In both flow regimes, feature maps are found to extract similar sets of flow structures such as braid shear-layers and shedding vortices. A Fourier analysis is conducted to investigate mechanisms of the CNN system for predicting wake flow in flow regimes with different wavenumber characteristics. It is found that a convolution layer in the CNN system integrates and transports wavenumber information from flow to predict the dynamics.
Characteristics of the CNN system for transporting input information including time histories of flow variables are analyzed by assessing contributions of each flow variable and time history to feature maps in the CNN system.
Structural similarities among feature maps in the CNN system are calculated to reveal the number of feature maps that contain similar flow structures. By reducing feature maps which contain similar flow structures, it is also able to successfully reduce the number of weights to learn in the CNN system without affecting prediction performance.\\
\end{abstract}

\maketitle
\section{Introduction}\label{sec:introduction}
The achievement of the state-of-the-art performance on a task of classifying images using a convolutional neural network (CNN)~\cite{krizhevsky2012imagenet} has boosted the usage of CNNs to tasks in computer vision~\cite{cirecsan2012multi,he2016identity,ren2015faster,chen20153d,xu2015show,ren2015faster}.
CNNs are reported to be good at extracting spatial features~\cite{masci2011stacked,zhao2016spectral,he2017mask,luo2016understanding} as it incorporates geometric knowledge of data into the network~\cite{lecun1989backpropagation}.
This attribute of CNNs has led to the success of CNNs to learn data with spatial features, such as images or videos. Opportunely, data of flow fields inherently contain spatial features, i.e., flow structures, governed by the Navier-Stokes equations. 
The existence of flow structures has been revealed in the numerous previous studies by experiments~\cite{freymuth1966transition,ruderich1986experimental,jaunet2017two}, simulations~\cite{waleffe1998three,wu2009direct,babucke2008dns,rockwood2018tracking,chong2015condensation,biferale2016coherent,francois2014three,faisst2003traveling}, and decomposition methods~\cite{sirovich1987turbulence, aubry1988dynamics, berkooz1993proper, nair2018networked,schmid2010dynamic,kutz2016dynamic,roy2015deconvolution,mezic2005spectral,mezic2013analysis}. Therefore, CNNs are expected to be utilizable for learning and predicting fluid dynamics.

Consequently, the recent developments in CNN architectures have yielded new methods to approximate flow at states such as geometries or Reynolds numbers that were not utilized during training~\cite{guo2016convolutional,miyanawala2017efficient,lee2017prediction,lee_you_2019,bhatnagar2019prediction}. For instance, \citet{lee_you_2019} developed a generative adversarial network (GAN), which is composed of multiple CNNs, to predict flow over a circular cylinder on two-dimensional slices at Reynolds numbers that were not utilized during training. They reported the practical usage of CNNs by predicting flow over a circular cylinder during several flow cycles. For more practical usage, the GAN has been employed to predict typhoon tracks~\cite{ruttgers2019prediction}. The capability of CNNs to approximate flow at states that were not utilized during training has provided potential for using CNNs to predict or model flow characteristics in practical problems.

However, despite the success in the application of CNN-based network systems to predict fluid dynamics, mechanisms of CNNs to learn fluid dynamics remain as black boxes. As a result, inevitable extensive parameter studies and trial-and-errors occur to develop a network. Understanding on mechanisms of CNNs to learn fluid dynamics would provide insights on developing CNN systems for physical problems, leading to reduced parameter studies and trial-and-errors during the development.

In the present study, a CNN system which consists of multiple CNNs with different resolution for predicting three-dimensional unsteady wake flow dynamics is developed, and mechanisms of the CNN system for predicting wake flow behind a circular cylinder are investigated in two flow regimes: the three-dimensional wake transition regime and the shear-layer transition regime. Information in feature maps in the CNN system is visualized to understand characteristics of flow structures that a feature map extracts in different flow regimes and to identify effects of stacking convolution layers in the CNN system. It is worth noting that a few studies have visualized feature maps to understand CNNs for classification problems~\cite{zeiler2014visualizing, yosinski2015understanding, simonyan2013deep}. In addition, a Fourier analysis is performed to investigate mechanisms of the CNN system to predict wake flow dynamics in different length scales. It is investigated that how the CNN system integrates and transports the input information for prediction of fluid dynamics. Based on an analysis of feature maps, it is attempted to reduce the number of feature maps in the CNN system.

The paper is organized as follows: methods associated with the CNN system and constructing flow datasets are explained in Section~\ref{sec:methods}. Mechanisms of the CNN system to learn wake flow and attempts to reduce the number of feature maps in the CNN system are discussed in Section~\ref{sec:results}, followed by concluding remarks in Section~\ref{sec:conclusion}.

\section{Methods}{\label{sec:methods}}
\subsection{Construction of datasets of unsteady flow fields}
\begin{figure}
  \centering
  \includegraphics[width=0.45\linewidth,trim={0.00cm 0.00cm 0.00cm 0.00cm},clip]{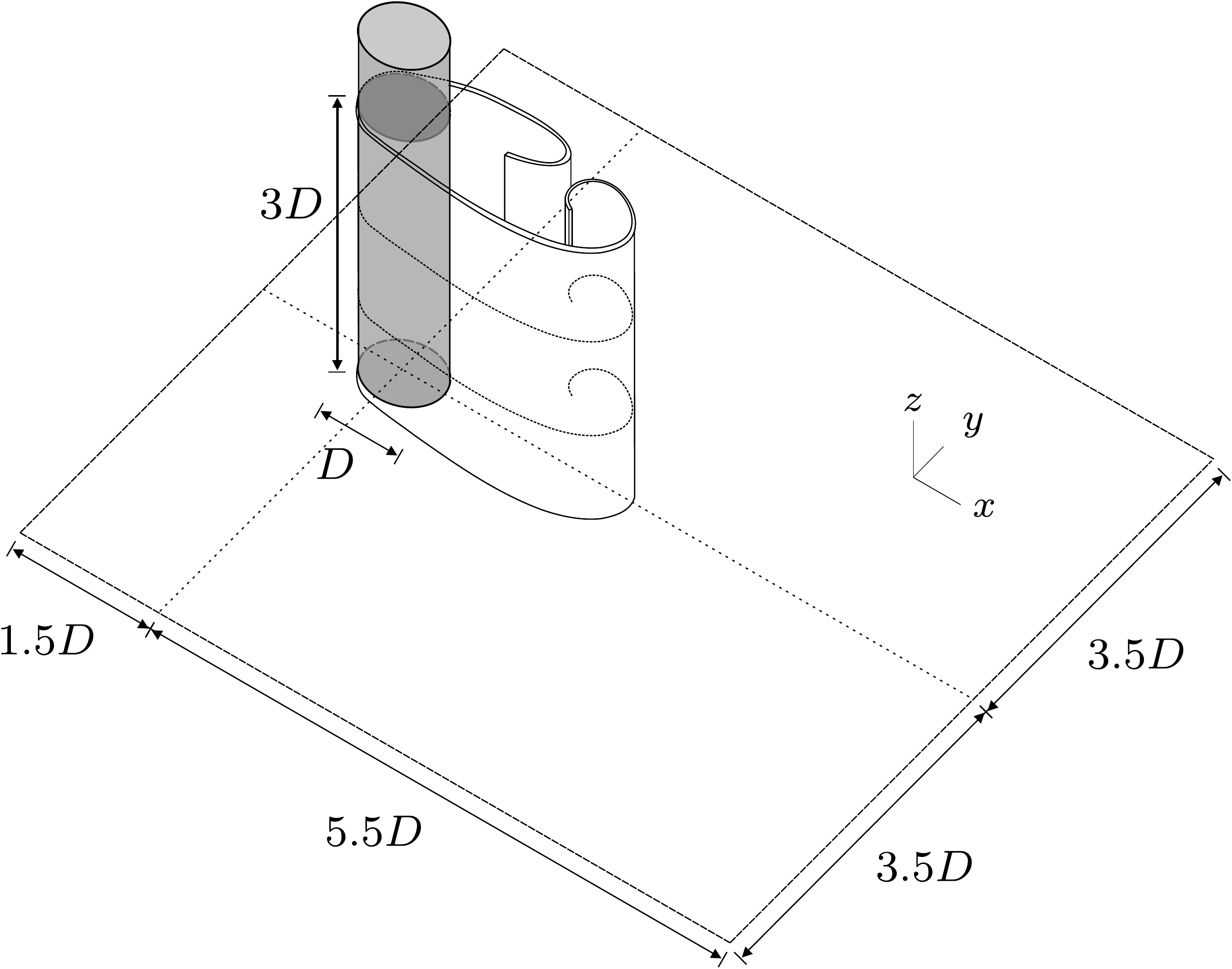}
  \caption{Schematic illustration of the domain for collecting data of flow over a circular cylinder.}
  \label{fig:flow}
\end{figure}
Unsteady flow over a circular cylinder is generated by conducting numerical simulations of the incompressible Navier-Stokes equations as described in detail in~\citet{lee_you_2019}. Here, main features of the constructed flow field datasets are summarized. Let the positive $x,y$, and $z$ directions be the streamwise, cross-stream, and spanwise directions, respectively, then a circular cylinder is located on the surface of $x^{2}+y^{2}=(D/2)^{2}$ along the spanwise direction, where $D$ is the cylinder diameter.
Flow fields in the three-dimensional wake transition regime at Reynolds numbers $Re_{D}$ $(=U_{\infty}D/\nu)$ of $300$, $400$, and $500$ and the shear-layer transition regime at Reynolds numbers of $1000$, $2000$, and $3900$, where $U_\infty$ is the free-stream velocity and $\nu$ is the kinematic viscosity of the fluid, are numerically simulated. 

The velocity components and the pressure are non-dimensionalized by the freestream velocity $U_{\infty}$ and $\rho U^{2}_{\infty}$ where $\rho$ is the density of the fluid, respectively.
Flow fields of $500$ time steps with an interval size $\delta t$ of 0.1 in a domain near the circular cylinder ($-1.5<x/D<5.5, -3.5<y/D<3.5$, and $0<z/D<3$) are collected (see Fig.~\ref{fig:flow}). The interval size $\delta t$ of 0.1 is the unit for the time-step interval size for training and prediction of the present CNN system and corresponds to 20 times of the time-step size for numerical simulations ($20\times \frac{U_\infty \Delta t}{D}$), where $\frac{U_\infty \Delta t}{D}$ is $0.005$.

Flow fields at $Re_{D}=300, 500, 1000$, and $2000$ are utilized to train the CNN system (see Section~\ref{sec:CNN_arc} for the architecture of the developed CNN system), while flow fields at $Re_{D}=400$ and $3900$, which are within and out of the range of the trained Reynolds numbers, are utilized to test the CNN system.
During training, a series of flow fields is randomly cropped to domains with the size of $0.875D \times 0.875D \times 1.5D$ and with grid resolution of $32 \times 32 \times 32$.
These randomly cropped flow fields are fed as an input of the CNN system during training.
On the other hand, three overlapping domains ($D_{1}, D_{2}$, and $D_{3}$) that decompose the collected flow fields in the spanwise direction as $D_{1}:0<z/D<1.5$, $D_{2}:0.75<z/D<2.25$, and $D_{3}:1.5<z/D<3.0$ are fed as the input of the CNN system during testing. Predicted results from the three domains are merged by using parts of domains as: $D1:0<z/D<1.125$, $D2:1.125<z/D<1.875$, and $D3:1.875<z/D<3.0$. Merging of predictions from each domain enables the network to minimize effects from boundaries caused by the use of paddings and also enables to extend the domain size for prediction by preventing an overload of memory of a graphics processing unit (GPU).

\subsection{CNN system for predicting three-dimensional unsteady wake flow}\label{sec:CNN_arc}
A CNN system for predicting three-dimensional unsteady wake flow is developed. The developed CNN system learns a mapping between flow fields in the past and a flow field at a future state. The CNN system shares a similar architecture with the CNN system developed by~\citet{lee_you_2019}, which is composed of two-dimensional convolution layers with zero padding to predict fluid flow on two-dimensional slices of flow fields. In the present study, the two-dimensional convolution layers are extended to three-dimensional convolution layers to investigate mechanisms of a CNN system to learn three-dimensional  nature fluid flow. Also, instead of the zero padding, the symmetric padding, which pads boundary values around feature maps, is utilized to prevent sharp changes of values near boundaries during convolution operations.

\begin{figure}
  \centering
  \subfigure[]{\includegraphics[width=0.9\linewidth,trim={0.00cm 0.00cm 0.00cm 0.00cm},clip]{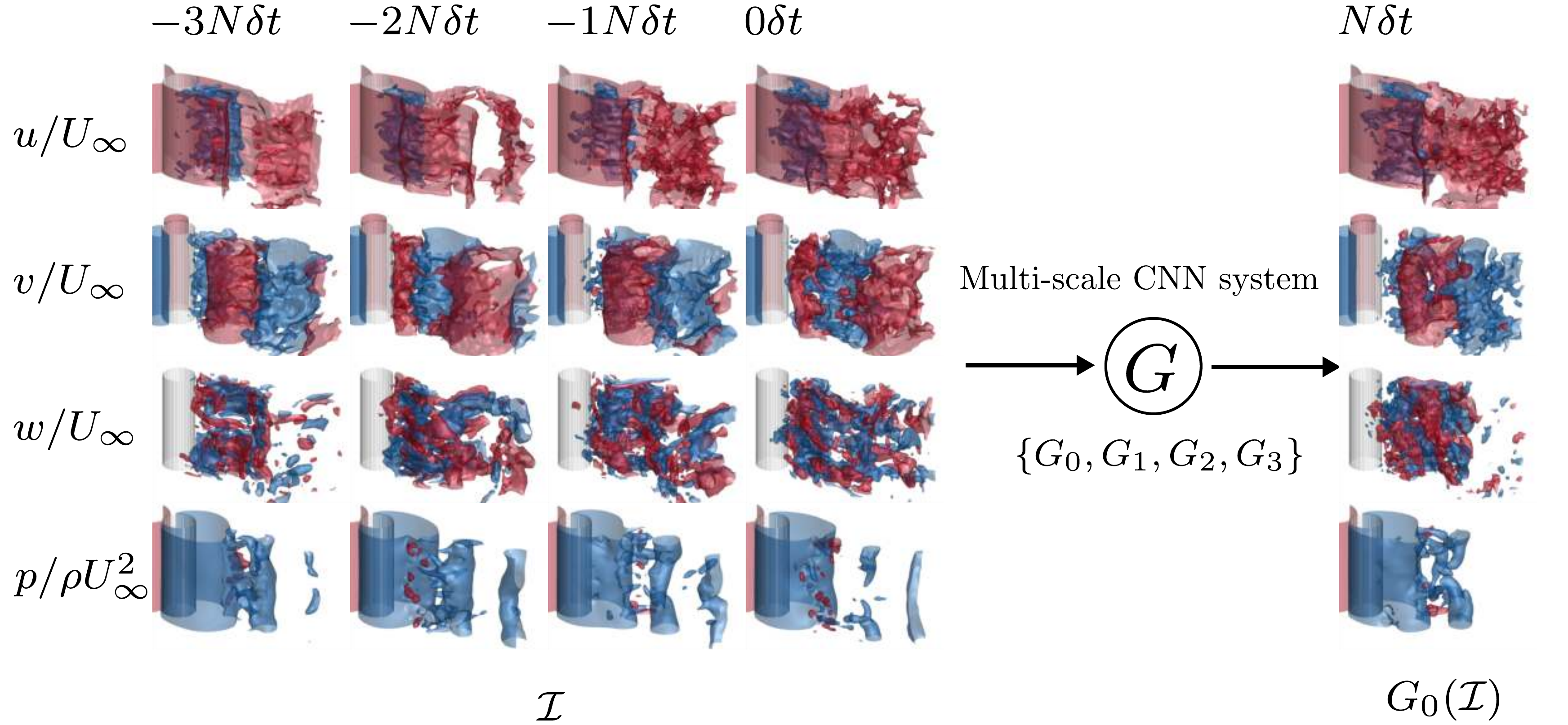}}
  \subfigure[]{\includegraphics[width=0.7\linewidth,trim={0.00cm 0.00cm 0.00cm 0.00cm},clip]{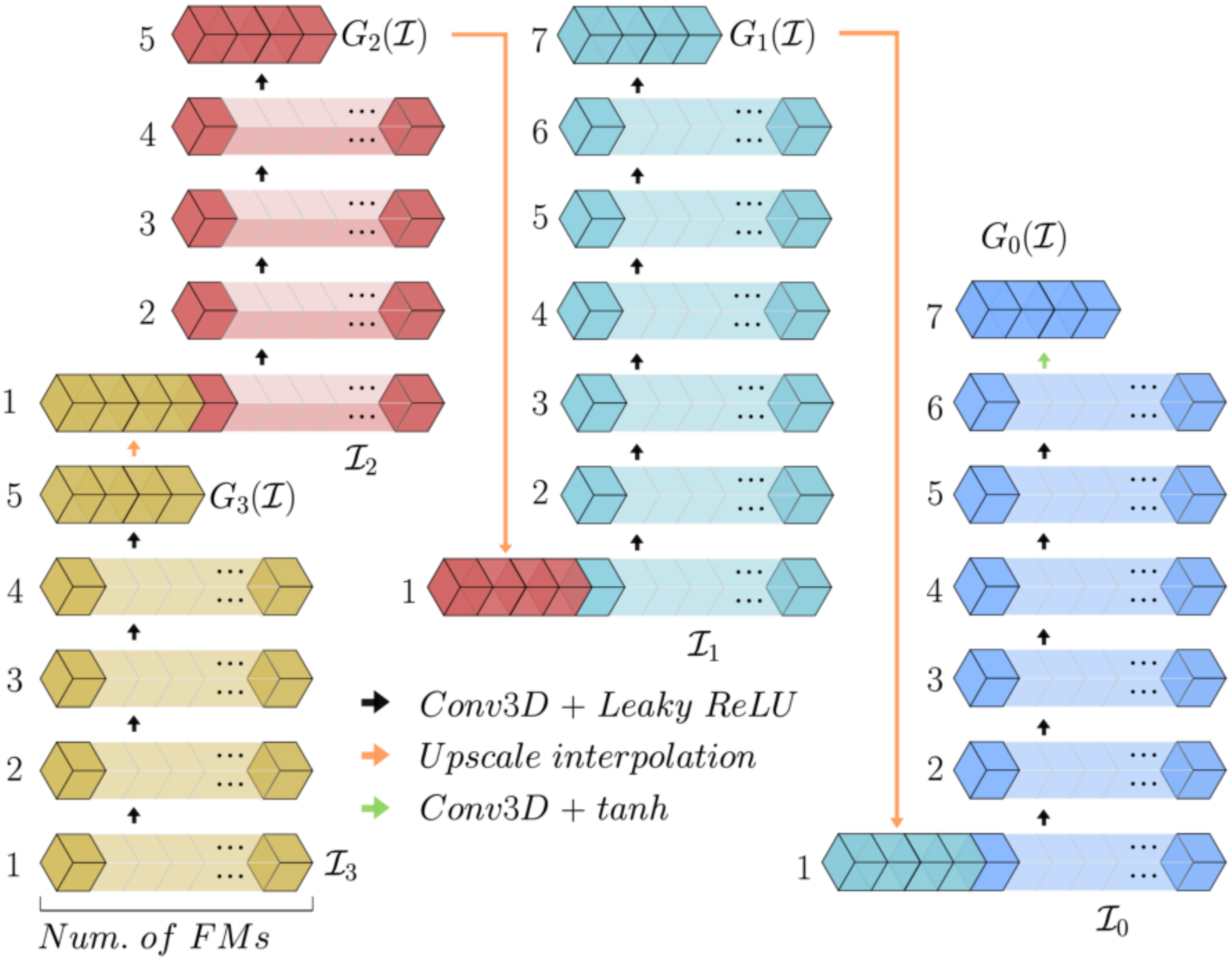}}
  \caption{Schematic illustration of the (a) input and output flow fields of the (b) developed multi-scale CNN system. Activation functions of $leaky$ $ReLU$ ($f(x) = max(0.2x,x)$) and $tanh$ ($f(x)=tanh(x)$) are utilized. The number of cubes represents the number of feature maps (Num. of FMs). Different colors of arrows indicate different types of connections. The merging of two blocks with different colors indicates a concatenation of feature maps. Layer numbers of feature maps are annotated for each generative CNN.}
  \label{fig:CNN}
\end{figure}

The developed CNN system is composed of a set of generative CNNs $\{G_{0}, G_{1}, G_{2}, G_{3}\}$ that are fed with input flow fields with different grid resolution. Due to this attribute, the developed CNN system can be considered as a multi-scale CNN system. 
The input flow fields contain information of velocity components ($u,v,w$) and pressure ($p$) at the past four sequential  times steps ($-3N\delta t$, $-2N\delta t$, $-1N\delta t$, $0\delta t$) with a constant time-step interval size $N\delta t$, where $N$ is a natural number (see Fig.~\ref{fig:CNN}(a) for a schematic illustration of the input and output flow fields for the CNN system). Let $X$, $Y$, and $Z$ be sizes of the domain in the streamwise, cross-stream, and spanwise directions, respectively, and $n_{x}$, $n_{y}$, and $n_{z}$ be the corresponding numbers of cells in the corresponding directions. Then, for an integer $s=0,1,2,$ and $3$, a generative CNN $G_{s}$ is fed with flow fields $\mathcal{I}_{s}$ with the domain size of $(X,Y,Z)$ and the number of cells of $(n_{x}/2^{s},n_{y}/2^{s},n_{z}/2^{s})$. Input flow fields $\mathcal{I}_{1}$, $\mathcal{I}_{2}$, and $\mathcal{I}_{3}$ are interpolated from the input flow field $\mathcal{I}_{0}$ with the finest grid resolution.

\begin{figure}
  \centering
  \includegraphics[width=0.60\linewidth,trim={0.0cm 0.0cm 0.0cm 0.0cm},clip]{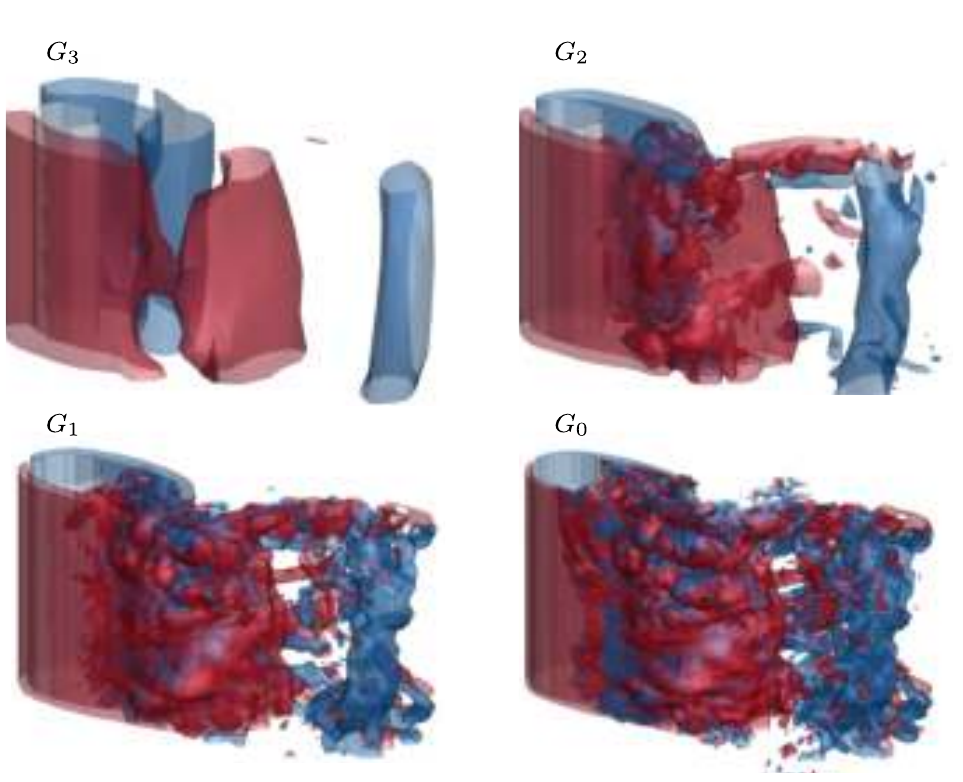}
  \caption{Isosurfaces of the instantaneous streamwise ($\omega_{x}D/U_{\infty}$) and spanwise ($\omega_{z}D/U_{\infty}$) vortices in the wake of a circular cylinder at $Re_{D}=3900$ predicted from $G_{0}$, $G_{1}$, $G_{2}$, and $G_{3}$ in the CNN system. Red-colored isosurfaces, $\omega_{x}D/U_{\infty}=\omega_{z}D/U_{\infty} = 2.0$; blue-colored isosurfaces, $\omega_{x}D/U_{\infty}=\omega_{z}D/U_{\infty} = -2.0$.}
  \label{fig:vor-scale}
\end{figure}

Generative CNNs $G_{1}$, $G_{2}$, and $G_{3}$ predict fields $\{G_{1}(\mathcal{I}), G_{2}(\mathcal{I}), G_{3}(\mathcal{I})\}$ of velocity components and pressure in coarse grid resolution and the information of the coarse prediction is transported to generative CNNs which are fed with flow fields with one-step higher grid resolution. Then, the generative CNN $G_{0}$ which is fed with flow fields with the finest grid resolution provides the final prediction $G_{0}(\mathcal{I})$ of velocity and pressure fields at a future state after the same time-step interval size $N\delta t$ as in the input flow fields. Details of connections in the utilized CNN system are described in Fig.~\ref{fig:CNN}(b), while an example of predictions in each generative CNN is shown in Fig.~\ref{fig:vor-scale}.
Details of the number of feature maps in the multi-scale CNN system are provided in Table~\ref{tab:CNN}. Three-dimensional convolution kernels with the size of $3 \times 3 \times 3$ are utilized for convolution operations on feature maps.
The present CNN system contains the number of weights which leads to a full utilization of the memory capacity of a single GPU (NVDIA TITAN XP) during testing.  
\begin{table}
\centering
\begin{tabular}{c|l}
\hline \hline
\multicolumn{1}{c|}{Generative CNN} & \multicolumn{1}{c}{Numbers of feature maps} \\ \hline
\multicolumn{1}{c|}{$G_{3}$}             & 16, 128, 256, 128, 4\\ 
\multicolumn{1}{c|}{$G_{2}$}             & 20, 128, 256, 128, 4\\ 
\multicolumn{1}{c|}{$G_{1}$}             & 20, 128, 256, 512, 256, 128, 4\\ 
\multicolumn{1}{c|}{$G_{0}$}             & 20, 128, 256, 512, 256, 128, 4\\ 
\hline \hline
\end{tabular}
\caption{Configuration of the present multi-scale CNN system. Convolution kernels with the size of $3\times 3 \times 3$ are utilized.}
\label{tab:CNN}
\end{table}

\begin{figure}
  \centering
  \includegraphics[width=0.60\linewidth,trim={0.0cm 0.0cm 0.0cm 0.0cm},clip]{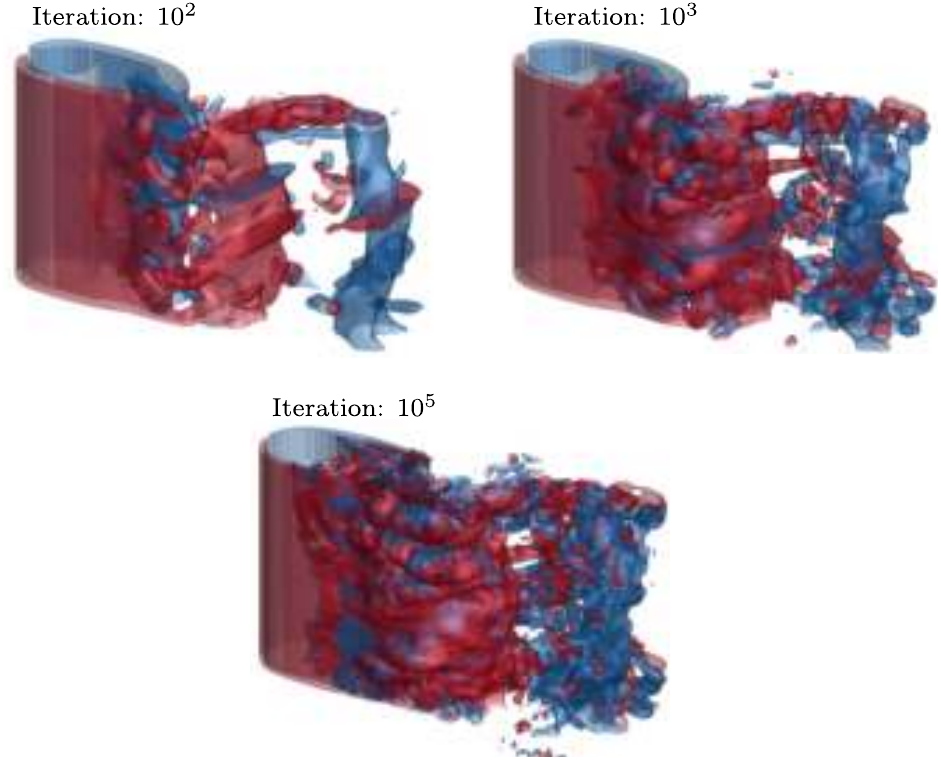}
  \caption{Isosurfaces of the instantaneous streamwise ($\omega_{x}D/U_{\infty}$) and spanwise ($\omega_{z}D/U_{\infty}$) vortices in the wake of a circular cylinder at $Re_{D}=3900$ predicted at $1\delta t$ from the CNN system trained with $10^{2}$, $10^{3}$, and $10^{5}$ iterations. Red-colored isosurfaces, $\omega_{x}D/U_{\infty}=\omega_{z}D/U_{\infty} = 2.0$; blue-colored isosurfaces, $\omega_{x}D/U_{\infty}=\omega_{z}D/U_{\infty} = -2.0$.}
  \label{fig:iter}
\end{figure}

The present study combines an $\mathcal{L}_{2}$ loss and physical losses of mass $\mathcal{L}_{c}$ and momentum $\mathcal{L}_{mom}$ conservation based on the Navier-Stokes equations, as advantages of providing knowledge of physics to a network have been reported~\cite{lee_you_2019, erichson2019physics}. The $\mathcal{L}_{2}$ loss is defined as follows:
\begin{equation}
\mathcal{L}_{2} = \sum_{s=0}^{3} ||G_{s}(\mathcal{I}) - \mathcal{G}_{s}(\mathcal{I})||_{2}^{2},
\label{eqn:loss_l2}
\end{equation}
where $G_{s}(\mathcal{I})$ and $\mathcal{G}_{s}(\mathcal{I})$ are the predicted flow field and the ground truth flow field at the same instance.
Let, $u$, $v$, $w$, and $p$ be the non-dimensionalized ground truth velocity components and pressure from $\mathcal{G}_{0}(\mathcal{I})$, while $\widetilde{u}$, $\widetilde{v}$, $\widetilde{w}$, and $\widetilde{p}$ be the predicted non-dimensionalized velocity components and pressure from $G_{0}(\mathcal{I})$ on a grid cell with the volume of $V_{i,j,k}$, where subscripts $i, j$, and $k$ are grid indices.
Then $\mathcal{L}_{c}$ and $\mathcal{L}_{mom}$ are calculated as follows:
\begin{equation}
\mathcal{L}_{c} = \sum_{i,j,k}^{n_{x},n_{y},n_{z}}\bigg|(\frac{\partial u}{\partial x}+ \frac{\partial v}{\partial y}+\frac{\partial w}{\partial z})-(\frac{\partial \widetilde{u}}{\partial x}+ \frac{\partial \widetilde{v}}{\partial y}+\frac{\partial \widetilde{w}}{\partial z})\bigg|V_{i,j,k},
\label{eq:loss_c}
\end{equation}
\begin{eqnarray*}
\mathcal{L}_{mom}^{x} = \{\frac{Du}{Dt} + \frac{\partial p}{\partial x} + \frac{1}{Re} (\frac{\partial^2 u}{\partial x \partial x}+\frac{\partial^2 u}{\partial y \partial y}+\frac{\partial^2 u}{\partial z \partial z})\} \nonumber\\
-\{\frac{D\widetilde{u}}{Dt} + \frac{\partial \widetilde{p}}{\partial x} + \frac{1}{Re} (\frac{\partial^2 \widetilde{u}}{\partial x \partial x}+\frac{\partial^2 \widetilde{u}}{\partial y \partial y}+\frac{\partial^2 \widetilde{u}}{\partial z \partial z})\}, \nonumber\\
\mathcal{L}_{mom}^{y} = \{\frac{Dv}{Dt} + \frac{\partial p}{\partial y} + \frac{1}{Re} (\frac{\partial^2 v}{\partial x \partial x}+\frac{\partial^2 v}{\partial y \partial y}+\frac{\partial^2 v}{\partial z \partial z})\}, \nonumber\\
-\{\frac{D\widetilde{v}}{Dt} + \frac{\partial \widetilde{p}}{\partial y} + \frac{1}{Re} (\frac{\partial^2 \widetilde{v}}{\partial x \partial x}+\frac{\partial^2 \widetilde{v}}{\partial y \partial y}+\frac{\partial^2 \widetilde{v}}{\partial z \partial z})\} \nonumber\\
\mathcal{L}_{mom}^{z} = \{\frac{Dw}{Dt} + \frac{\partial p}{\partial z} + \frac{1}{Re} (\frac{\partial^2 w}{\partial x \partial x}+\frac{\partial^2 w}{\partial y \partial y}+\frac{\partial^2 w}{\partial z \partial z})\} \nonumber\\
-\{\frac{D\widetilde{w}}{Dt} + \frac{\partial \widetilde{p}}{\partial z} + \frac{1}{Re} (\frac{\partial^2 \widetilde{w}}{\partial x \partial x}+\frac{\partial^2 \widetilde{w}}{\partial y \partial y}+\frac{\partial^2 \widetilde{w}}{\partial z \partial z})\}, \nonumber
\end{eqnarray*}
and
\begin{equation}
\mathcal{L}_{mom} = \sum_{i,j,k}^{n_{x},n_{y},n_{z}}\bigg| \mathcal{L}_{mom}^{x} + \mathcal{L}_{mom}^{y} + \mathcal{L}_{mom}^{z}\bigg|V_{i,j,k},
\label{eq:loss_mom}
\end{equation}
where temporal and spatial derivatives are calculated using the first-order backward scheme and second-order central difference schemes, respectively.
The total loss is calculated as follows:
\begin{eqnarray}
\mathcal{L}_{generator} = \frac{1}{\lambda_{\sum}}(\lambda_{l2} \mathcal{L}_{2} + \lambda_{c} \mathcal{L}_{c} + \lambda_{m} \mathcal{L}_{mom}),
\label{eqn: L_gen}
\end{eqnarray}
where $\lambda_{\sum} =  \lambda_{l2} + \lambda_{c} + \lambda_{m}$.
Contributions of each loss function can be controlled by tuning coefficients of losses $\lambda_{l2}$, $\lambda_{c}$, and $\lambda_{m}$.
The present study utilizes $\lambda_{l2}=1.0$, $\lambda_{c}=1.0$, and $\lambda_{m}=0.1$ to make contributions of losses $\mathcal{L}_{2}$, $\mathcal{L}_{c}$, and $\mathcal{L}_{mom}$ in a comparable level.
The CNN system is trained up to $6\times 10^{5}$ iterations. The trained CNN system with the smallest $L_{\infty}$ norm on test data, flow at $Re_{D}=400$ and $3900$, is selected for further analyses. The CNN system is found to learn large-scale structures first, while it learns smaller scale structures as the iteration number increases (see Fig.~\ref{fig:iter}). 
\section{Results and discussion}\label{sec:results}
\subsection{Mechanisms of the CNN system to learn wake flow}\label{subsec:Re}
\subsubsection{Prediction at $Re_{D}=400$ and $Re_{D}=3900$}
\begin{figure*}
  \centering
  \subfigure[]{\includegraphics[width=0.90\linewidth,trim={0.0cm 0.0cm 0.0cm 0.0cm},clip]{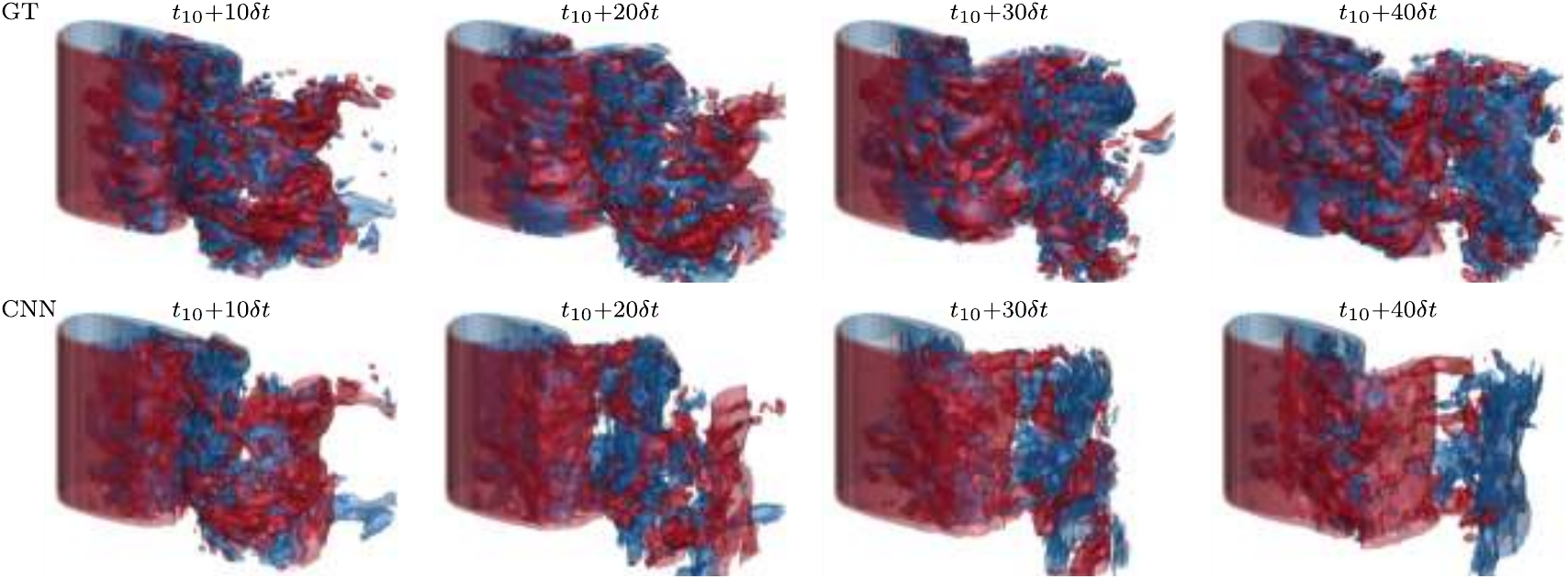}}
  \subfigure[]{\includegraphics[width=0.90\linewidth,trim={0.0cm 0.0cm 0.0cm 0.0cm},clip]{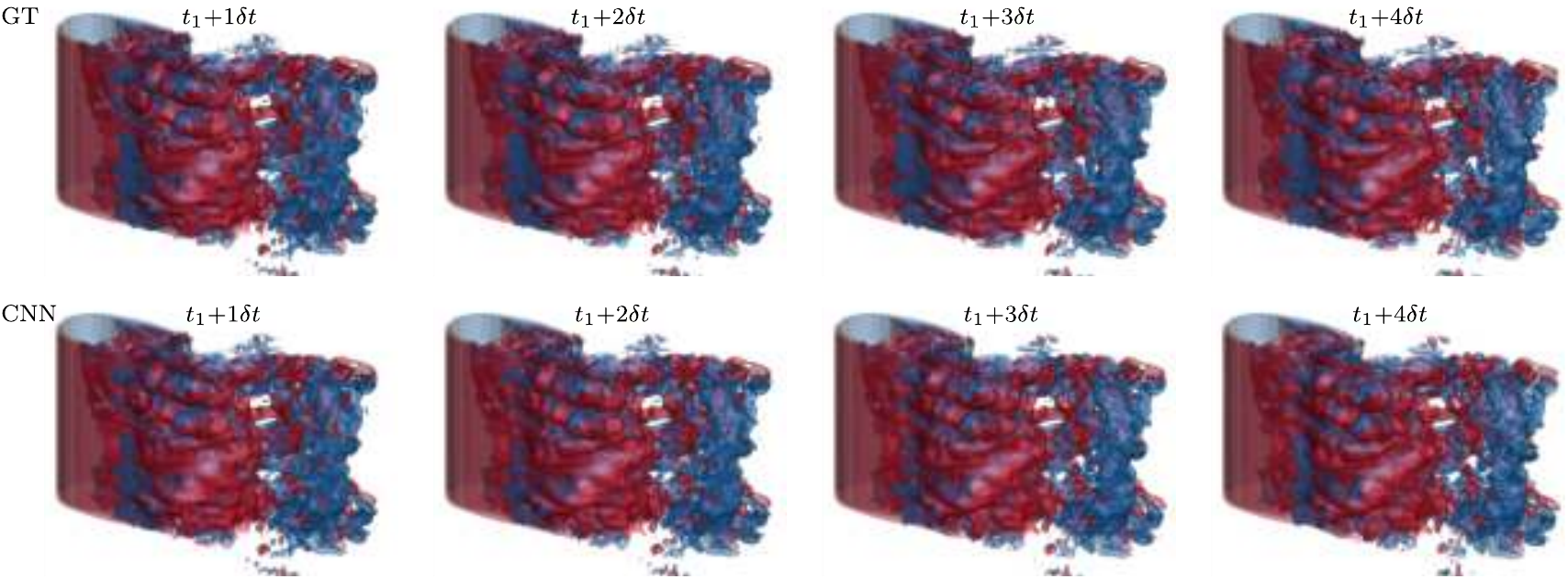}}
  \caption{Isosurfaces of the instantaneous streamwise ($\omega_{x}D/U_{\infty}$) and spanwise ($\omega_{z}D/U_{\infty}$) vortices in the wake of a circular cylinder at $Re_{D}=3900$ calculated from ground truth flow fields (GT), and flow fields predicted by the CNN system (CNN) trained with time-step interval sizes of (a) $10\delta t$ and (b) $1\delta t$. Red-colored isosurfaces, $\omega_{x}D/U_{\infty}=\omega_{z}D/U_{\infty} = 2.0$; blue-colored isosurfaces, $\omega_{x}D/U_{\infty}=\omega_{z}D/U_{\infty} = -2.0$.}
 \label{fig:iso-vor-T}
\end{figure*}

\begin{figure*}
  \centering
 \subfigure[]{\includegraphics[width=0.4\linewidth,trim={0.60cm 0.60cm 0.60cm 0.60cm},clip]{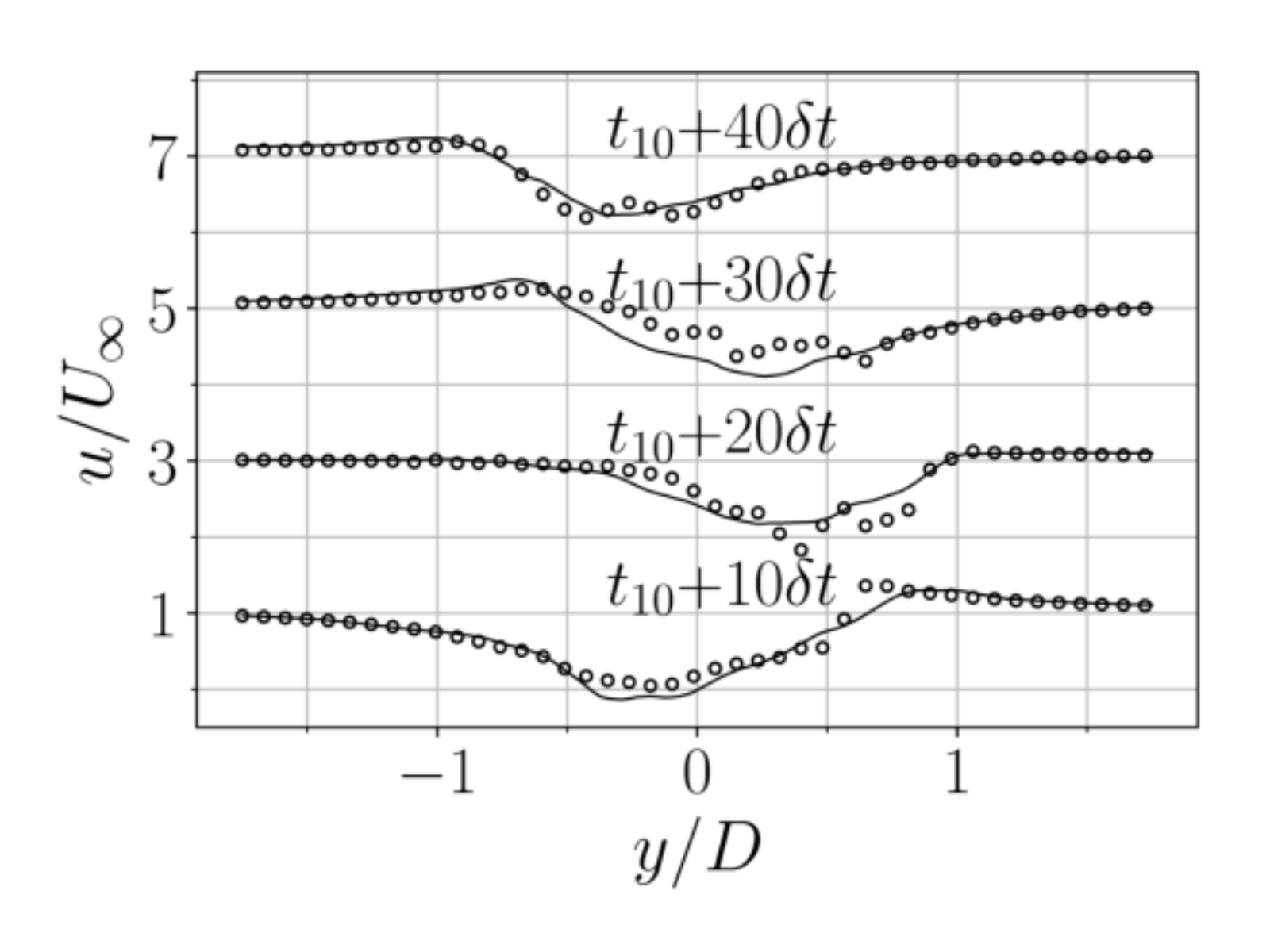}}
 \subfigure[]{\includegraphics[width=0.4\linewidth,trim={0.60cm 0.60cm 0.60cm 0.60cm},clip]{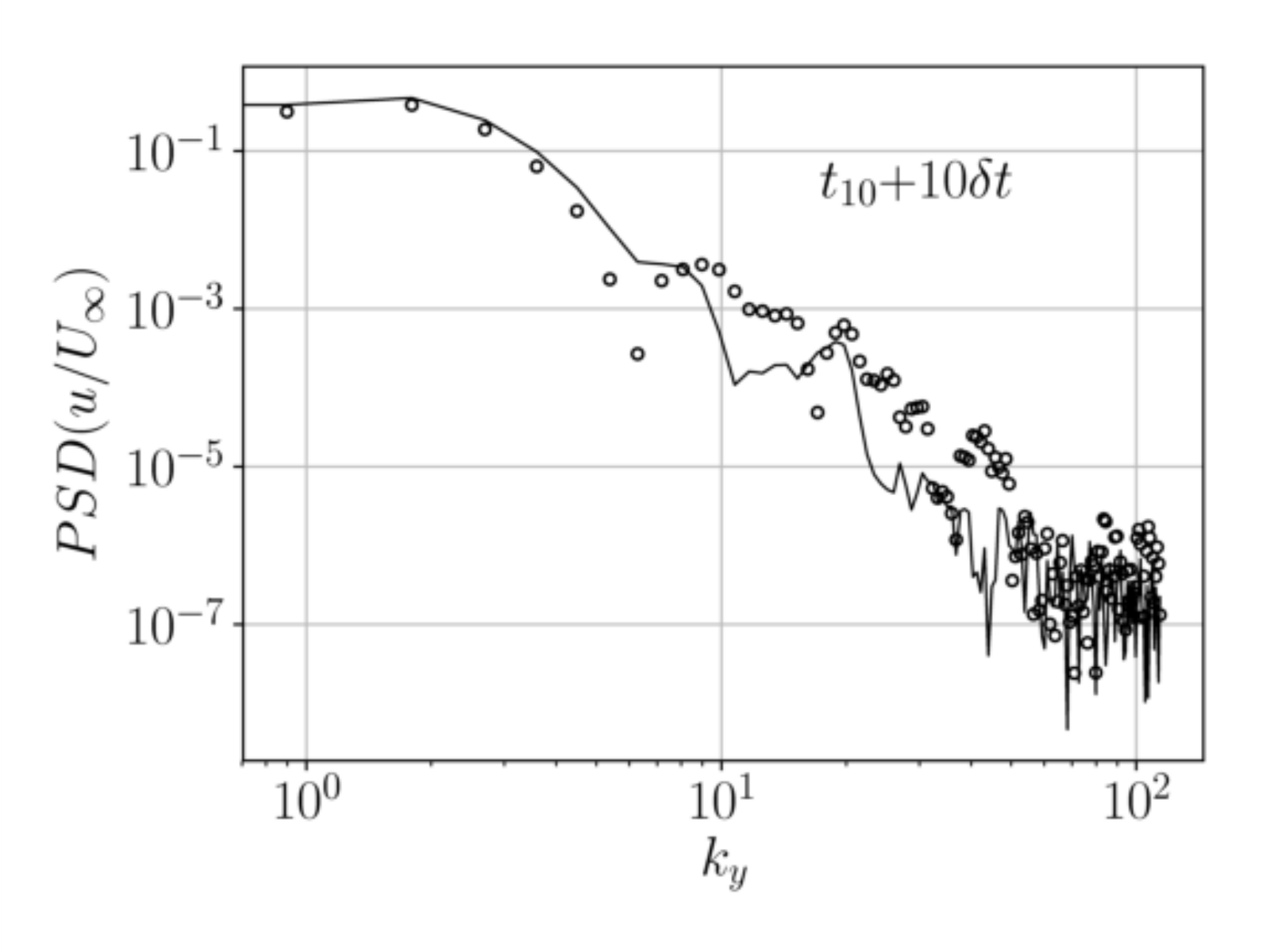}}

 \subfigure[]{\includegraphics[width=0.4\linewidth,trim={0.60cm 0.60cm 0.60cm 0.60cm},clip]{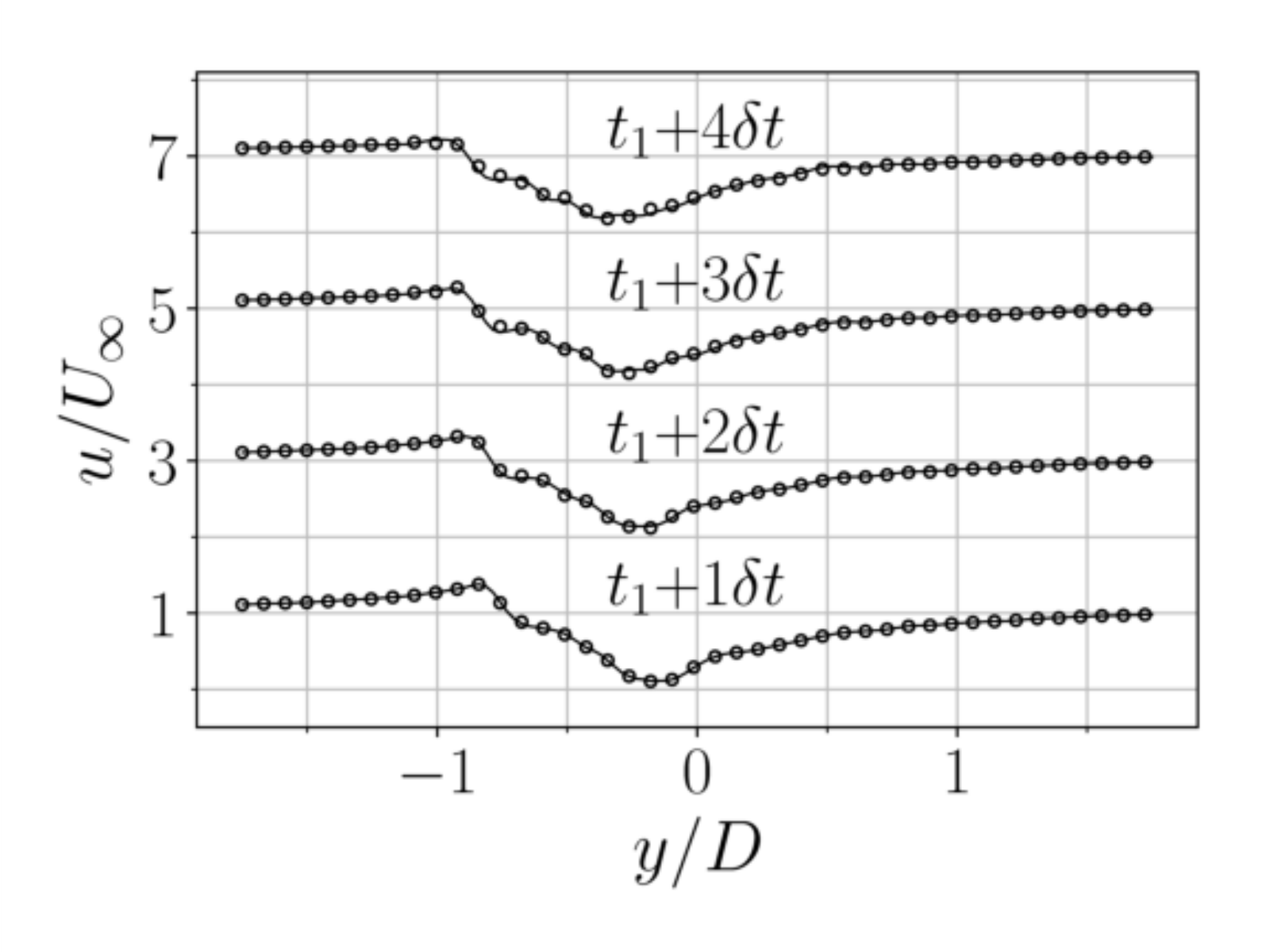}}
 \subfigure[]{\includegraphics[width=0.4\linewidth,trim={0.60cm 0.60cm 0.60cm 0.60cm},clip]{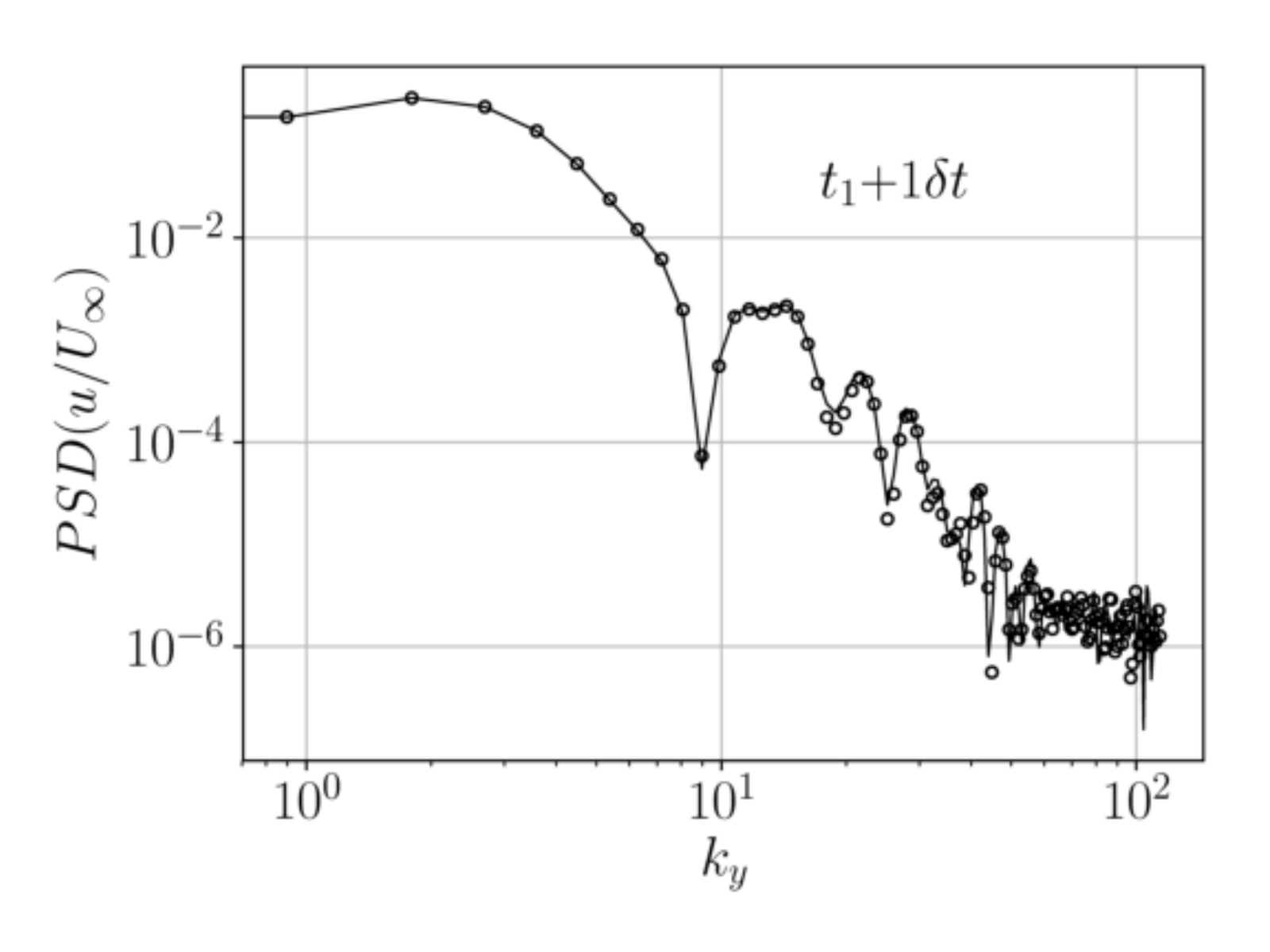}}
  \caption{Profiles ((a) and (c)) and PSDs ((b) and (d)) of the instantaneous streamwise velocity along a line at $x/D=2.0$ and $z/D=1.5$ at $Re_{D}=3900$ with time-step interval sizes of $10\delta t$ ((a) and (b)) and $1\delta t$ ((c) and (d)). Solid lines, from predicted flow fields using the CNN system; and circles, from ground truth flow fields. Profiles of the streamwise velocity after $N \times 1\delta t$ and $N \times 10\delta t$ are shifted by $2.0 \times (N-1)$ along the vertical axis, respectively, where $N=1,2,3,4$.}
 \label{fig:line-T}
\end{figure*}

Flow fields in the three-dimensional wake transition regime ($Re_{D}=400$) and the shear-layer transition regime ($Re_{D}=3900$) are predicted by employing the CNN system trained with flow fields at Reynolds numbers of $Re_{D}=300, 500, 1000$, and $2000$.
Isosurfaces of the spanwise ($\omega_{z}D/U_{\infty}$) and streamwise ($\omega_{x}D/U_{\infty}$) vortices at $Re_{D}=3900$ calculated from prediction results of the CNN system trained with time-step interval sizes of $1\delta t$ and $10\delta t$ are compared with vortices calculated from ground truth flow fields (Fig.~\ref{fig:iso-vor-T}). Note that flow fields at time steps longer than $t_{1} + 1\delta t$ and $t_{10}+10\delta t$ are recursively predicted by feeding predictions in the previous time steps as the input of the network system, where $t_{1}$ and $t_{10}$ are reference time steps for flow fields with the time-step interval sizes of $1\delta t$ and $10\delta t$, respectively.
Large-scale oscillations of the separated shear-layers and the Karman vortex street are observed to be predicted in the flow fields generated by the CNN system trained with the time-step interval size of $10\delta t$ (see Fig.~\ref{fig:iso-vor-T}(a)).
Large-scale fluid motions are more quantitatively observable in the profiles of the streamwise velocity along the cross-stream direction at the location $x/D=2.0$ in the cylinder wake (see Fig.~\ref{fig:line-T}(a)).
The overall trends of profiles of the streamwise velocity at time steps with the interval size of $10\delta t$ are favorably  predicted by the present CNN system.

However, the discrepancy between the ground truth and predicted profiles of the streamwise velocity is observed in the small-scale spatial fluid motions, which implies that the network system trained with a large time-step interval size (e.g., $10\delta t$) is likely to miss small-scale flow structures.
As a consequence, power spectral densities (PSDs) of high wavenumber contents ($k_{y}>10$) in profiles of the streamwise velocity predicted by the CNN system trained with the time-step interval size of $10\delta t$ are found to be underpredicted (see Fig.~\ref{fig:line-T}(b)).
Also, lack of small-scale vortices in the prediction of the CNN system trained with the time-step interval size of $10\delta t$ is identifiable in Fig.~\ref{fig:iso-vor-T}(a). 
The loss of high wavenumber contents in flow is partly due to the incapability of learning small-scale flow motions in input flow fields of which time-scales are smaller than the time-step interval size for training, as discussed by~\citet{lee_you_2019} with a time-scale analysis.

In contrast to the CNN system trained with the time-step interval size of $10\delta t$, the CNN system trained with the time-step interval size of $1\delta t$ shows favorable capability to capture small-scale flow structures.
Profiles of the streamwise velocity and PSDs of high wavenumber contents ($k_{y}>10$) extracted from flow fields predicted by the CNN system trained with the time-step interval size of $1\delta t$ are found to agree well with the ground truth results (see Figs.~\ref{fig:line-T}(c)~and~(d)).
In addition, small-scale vortices are identifiable in the prediction results of the spanwise and streamwise vortices (see Fig.~\ref{fig:iso-vor-T}(b)).
As the CNN system trained with the time-step interval size of $1\delta t$ is found to be capable of predicting small-scale flow structures in flow at $Re_{D}=3900$, further investigations of fluid dynamics and mechanisms of the CNN system are conducted using the results from the CNN system trained with the time-step interval size of $1\delta t$. The reference time step $t_{1}$ is omitted in the further discussion.


\begin{figure*}
  \centering
  \includegraphics[width=0.95\linewidth,trim={0.0cm 0.0cm 0.0cm 0.0cm},clip]{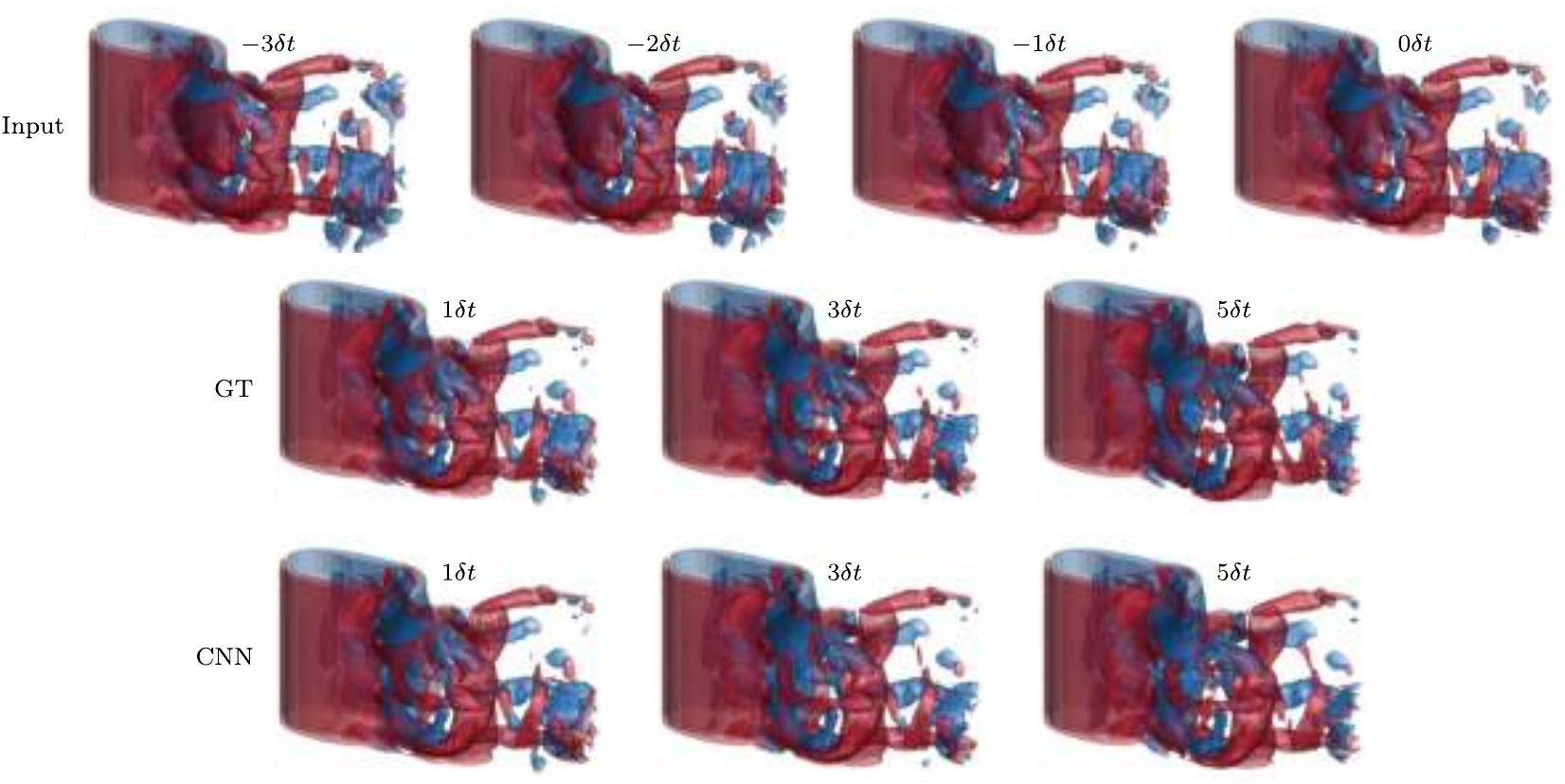}
  \caption{Isosurfaces of the instantaneous streamwise ($\omega_{x}D/U_{\infty}$) and spanwise ($\omega_{z}D/U_{\infty}$) vortices in the wake of a circular cylinder at $Re_{D}=400$ calculated from input flow fields (Input, $-3 \delta t$ to $0\delta t$), ground truth flow fields (GT, $1\delta t$, $3\delta t$, $5\delta t$), and flow fields predicted by  the CNN system (CNN, $1\delta t$, $3\delta t$, $5\delta t$). Red-colored isosurfaces, $\omega_{x}D/U_{\infty}=\omega_{z}D/U_{\infty} = 2.0$; blue-colored isosurfaces, $\omega_{x}D/U_{\infty}=\omega_{z}D/U_{\infty} = -2.0$.}
 \label{fig:iso-vor-400}
\end{figure*}

\begin{figure*}
  \centering
  \includegraphics[width=0.95\linewidth,trim={0.0cm 0.0cm 0.0cm 0.0cm},clip]{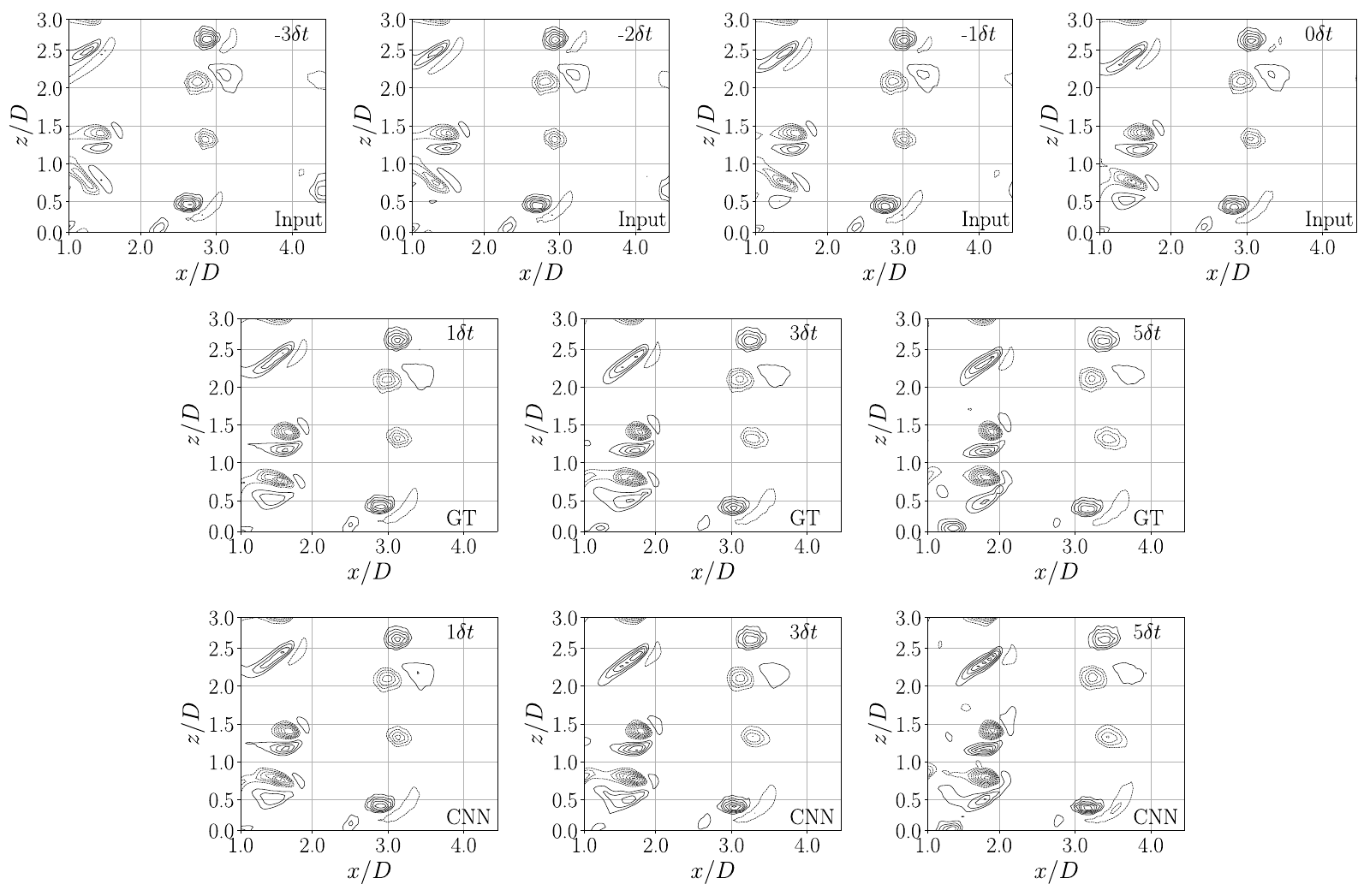}
  \caption{Instantaneous cross-stream vorticity in the wake of cylinder flow at $Re_{D}=400$ calculated from four input flow fields (Input, $-3\delta t$ to $0\delta t$), ground truth flow fields (GT, $1\delta t$, $3\delta t$, $5\delta t$), and  flow fields predicted by the CNN system (CNN, $1\delta t$, $3\delta t$, $5\delta t$). Slices are extracted on an $x$-$z$ plane at $y/D=0$. 20 contour levels from -10.0 to 10.0 are shown. Solid lines and dashed lines indicate positive and negative contour levels, respectively.}
  \label{fig:vor-sym-400}
\end{figure*}

\begin{figure}
  \centering
  \subfigure[]{\includegraphics[width=0.48\linewidth,trim={0.0cm 0.0cm 0.0cm 0.0cm},clip]{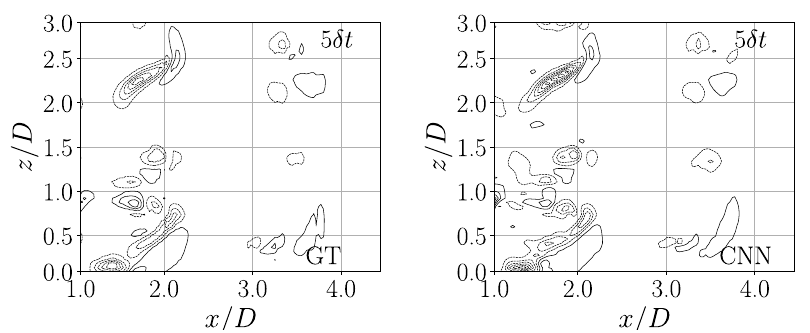}}
  \subfigure[]{\includegraphics[width=0.48\linewidth,trim={0.0cm 0.0cm 0.0cm 0.0cm},clip]{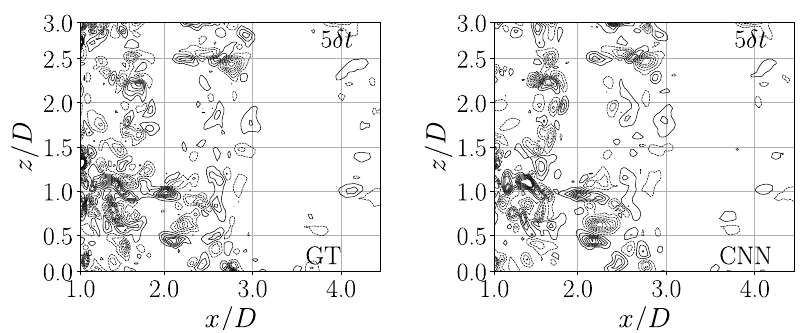}}

  \subfigure[]{\includegraphics[width=0.48\linewidth,trim={0.0cm 0.0cm 0.0cm 0.0cm},clip]{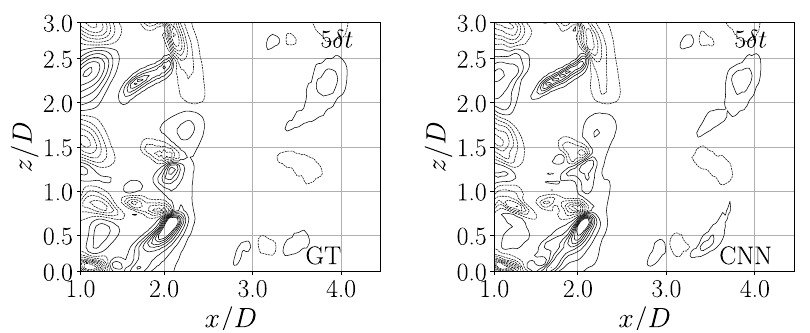}}
  \subfigure[]{\includegraphics[width=0.48\linewidth,trim={0.0cm 0.0cm 0.0cm 0.0cm},clip]{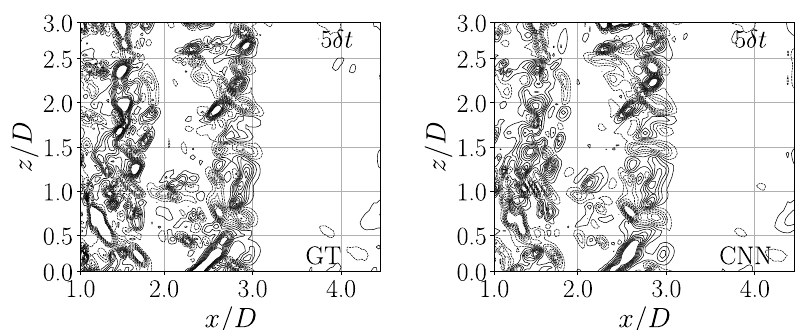}}

  \subfigure[]{\includegraphics[width=0.48\linewidth,trim={0.0cm 0.0cm 0.0cm 0.0cm},clip]{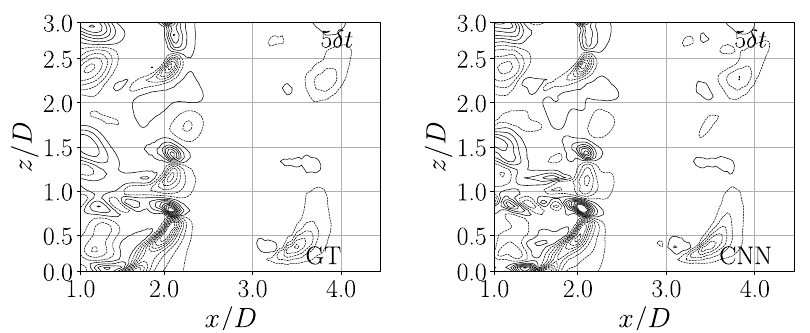}}
  \subfigure[]{\includegraphics[width=0.48\linewidth,trim={0.0cm 0.0cm 0.0cm 0.0cm},clip]{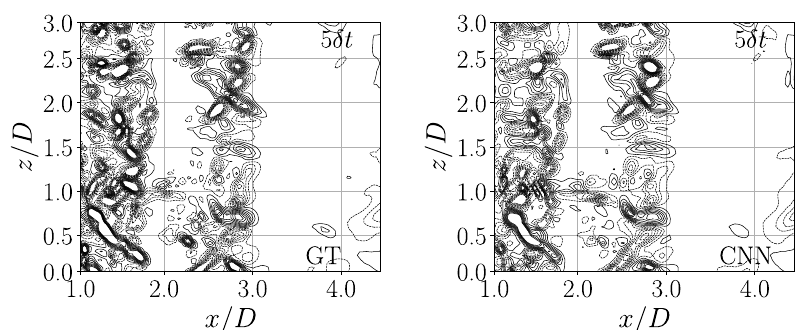}}
  \caption{Contours of the cross-stream vortex stretching ($\omega_{y} \frac{\partial v}{\partial y}D^{2}/U^{2}_{\infty}$) term ((a) and (b)), tilting term in the streamwise direction ($\omega_{x} \frac{\partial v}{\partial x}D^{2}/U^{2}_{\infty}$)  ((c) and (d)), and tilting term in the spanwise direction ($\omega_{z} \frac{\partial v}{\partial z}D^{2}/U^{2}_{\infty}$) ((e) and (f)) in the wake of cylinder flow at $Re_{D}=400$ ((a), (c), and (e)) and $3900$ ((b), (d), and (f)) calculated from the ground truth (GT) and the predicted (CNN) flow fields. Slices are extracted on the $x-z$ plane at $y/D=0$. 20 contour levels from -10.0 to 10.0 are shown. Solid lines and dashed lines indicate positive and negative contour levels, respectively.}
  \label{fig:vor-st}
\end{figure}

Isosurfaces of the spanwise and streamwise vortices calculated from the ground truth and predicted flow fields with the time-step interval size of $1\delta t$ at the Reynolds number of $Re_{D}=400$ are shown in Fig.~\ref{fig:iso-vor-400}. Note that the flow fields at $5\delta t$ are recursively predicted by only using predictions in the previous time steps. Flow structures in the three-dimensional wake transition regime are observed to be well predicted using the CNN system.
To take a closer look at the predicted flow dynamics at $Re_{D}=400$, slices on the symmetry plane ($x$-$z$ plane at $y/D=0$) with contours of the cross-stream vorticity are shown in Fig.~\ref{fig:vor-sym-400}.
Convection of vortices is clearly observed by comparing flow fields at two different time steps, for instance, at $1 \delta t$ and $5 \delta t$. Vortices located near $x/D = 1.5$ and $3.0$ at $1\delta t$ are convected to downstream locations at $5\delta t$ with a distance of approximately $0.25D$.
However, predicted flow at $5\delta t$ shows a small discrepancy of vortical structures at locations of $x/D<2.5$, where stretching and tilting of vortices are expected to abundantly occur in a vortex formation region.
The abundant occurrence of vortex stretching ($\omega_{y} \frac{\partial v}{\partial y}D^{2}/U^{2}_{\infty}$) and tilting ($\omega_{x} \frac{\partial v}{\partial x}D^{2}/U^{2}_{\infty}$, $\omega_{z} \frac{\partial v}{\partial z}D^{2}/U^{2}_{\infty}$) of the cross-stream vorticity in a vortex formation region can be observed in Fig.~\ref{fig:vor-st}.
Also, similarly to the discrepancy in fields of the cross-stream vorticity, notable discrepancy in vortex stretching and tilting occurs at locations of $x/D<2.5$.
For flow in the shear-layer transition regime, the discrepancy is expected to be larger, due to more abundant occurrence of stretching and tilting of vortices, compared to flow in the three-dimensional wake transition regime (see Fig.~\ref{fig:vor-st}).
\begin{figure}
  \centering
  \subfigure[]{\includegraphics[width=0.35\linewidth,trim={0.60cm 0.60cm 0.60cm 0.60cm},clip]{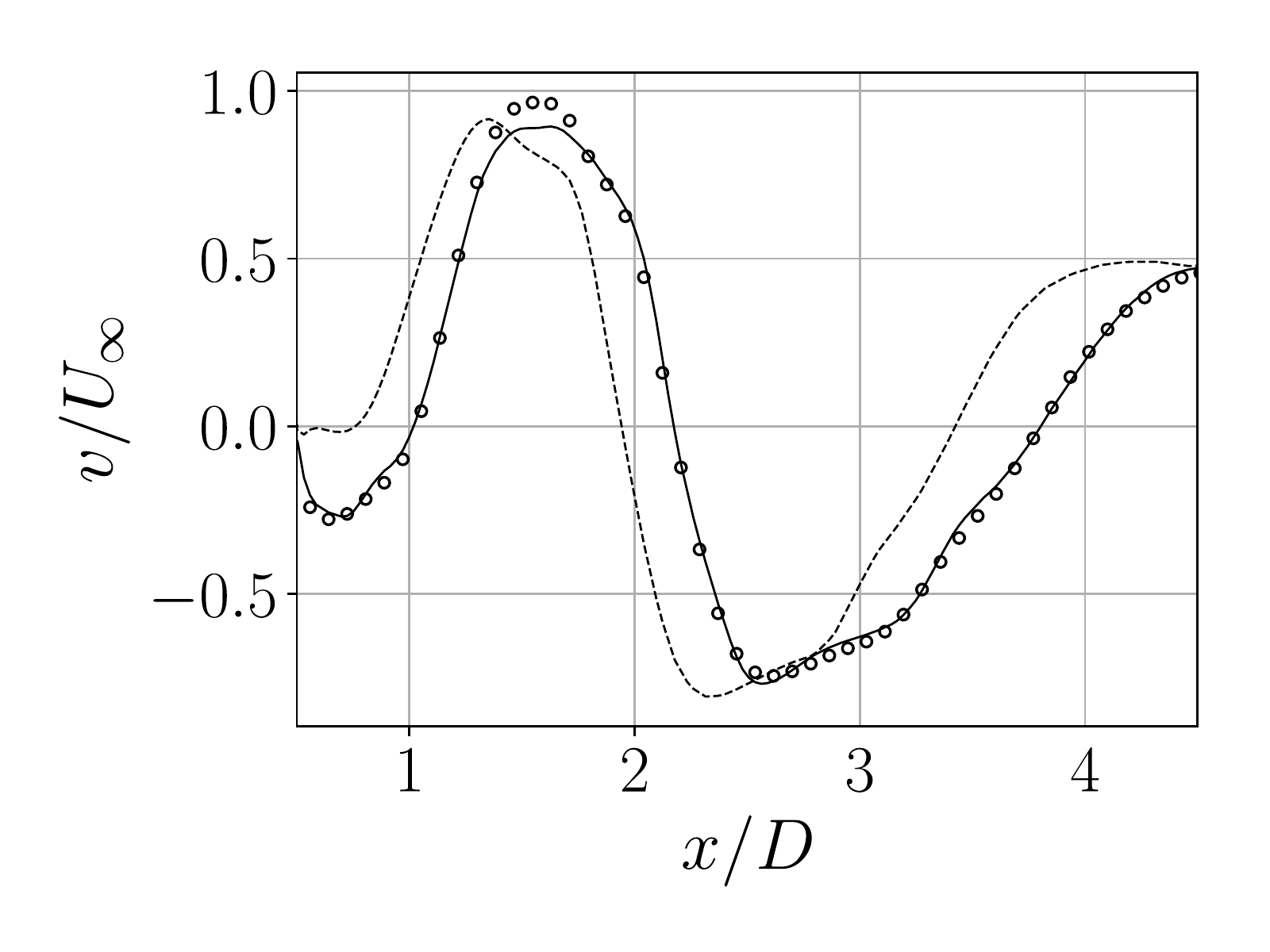}}
  \subfigure[]{\includegraphics[width=0.35\linewidth,trim={0.60cm 0.60cm 0.60cm 0.60cm},clip]{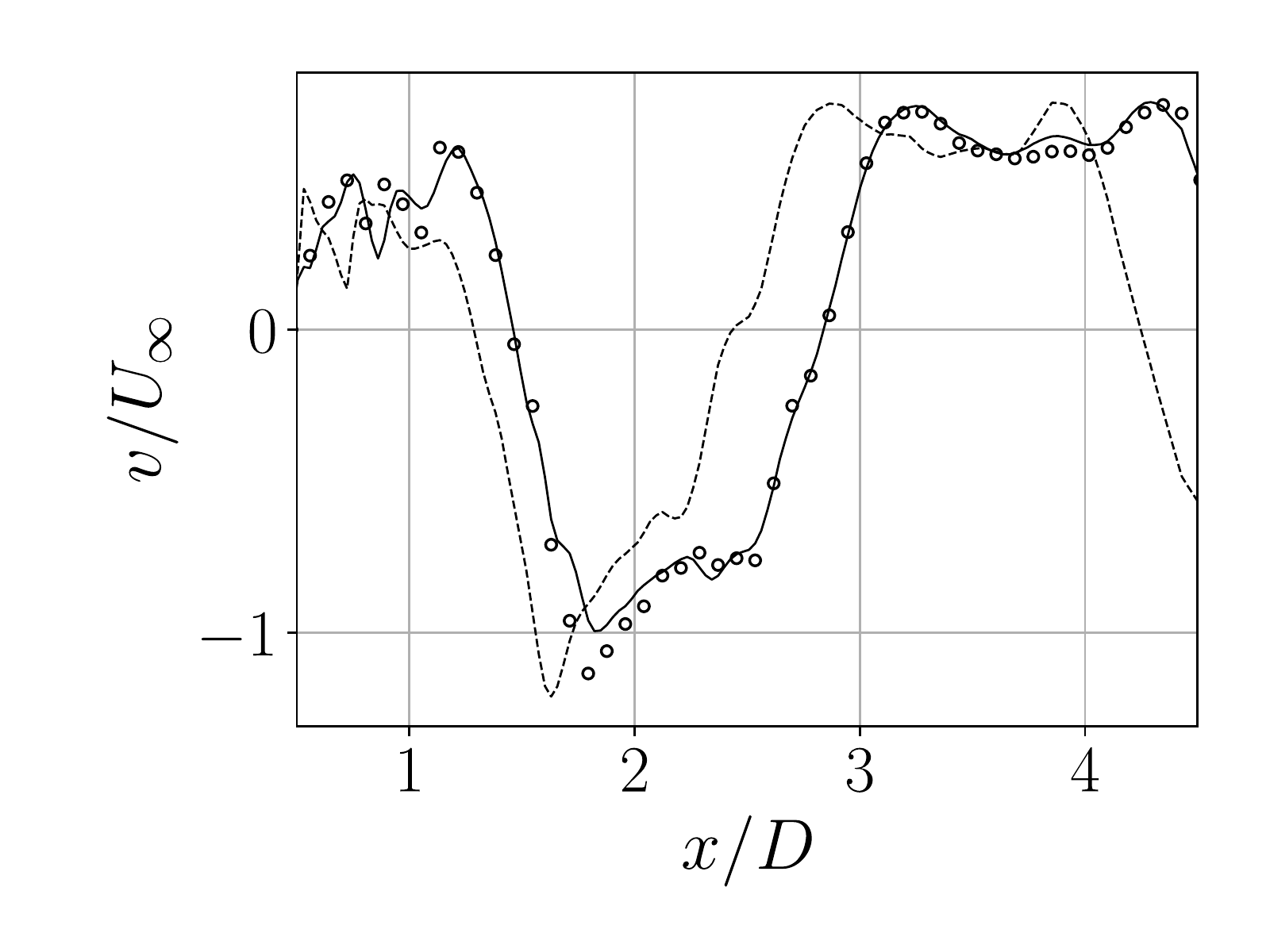}}

  \subfigure[]{\includegraphics[width=0.35\linewidth,trim={0.60cm 0.60cm 0.60cm 0.60cm},clip]{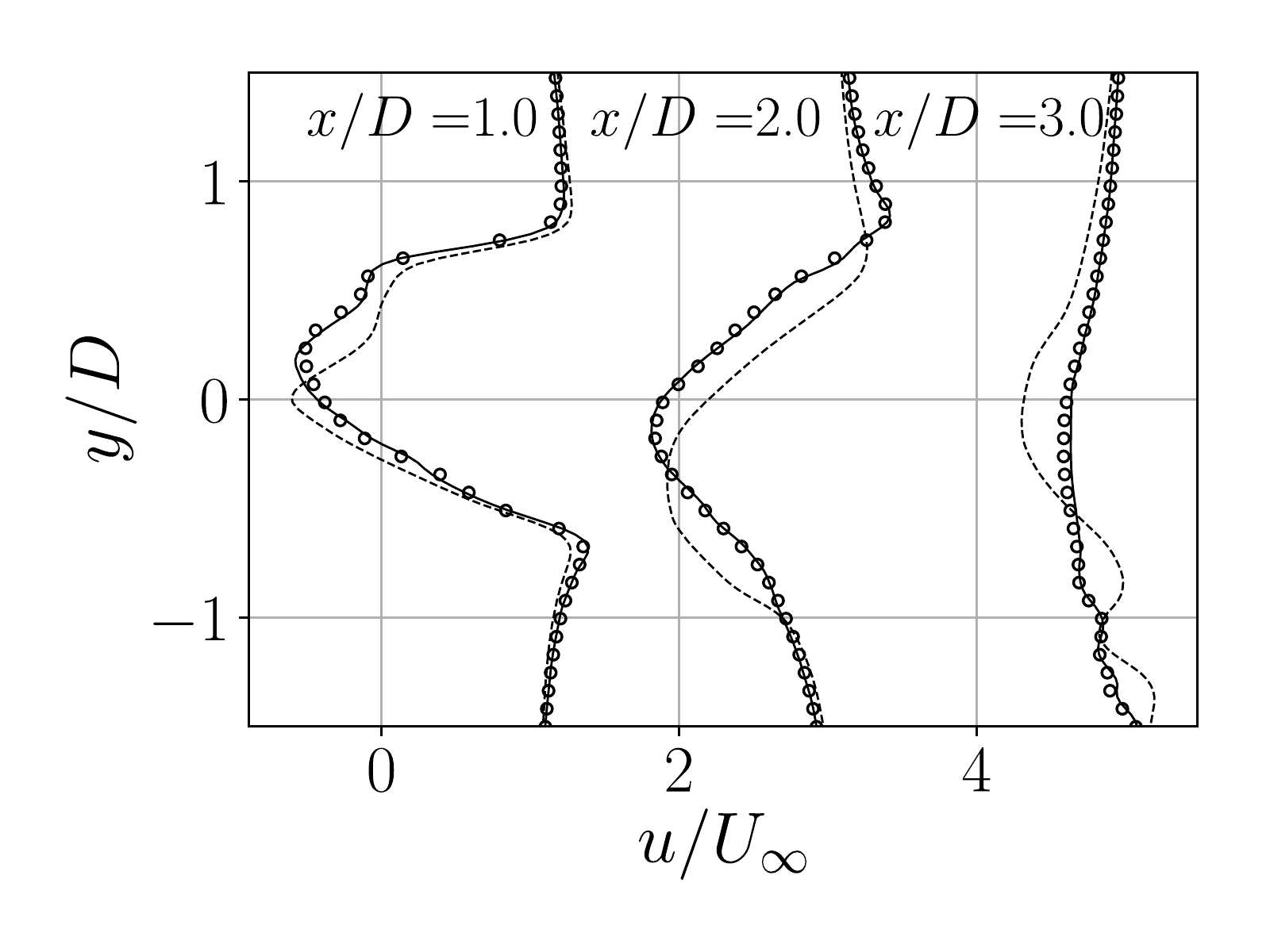}}
  \subfigure[]{\includegraphics[width=0.35\linewidth,trim={0.60cm 0.60cm 0.60cm 0.60cm},clip]{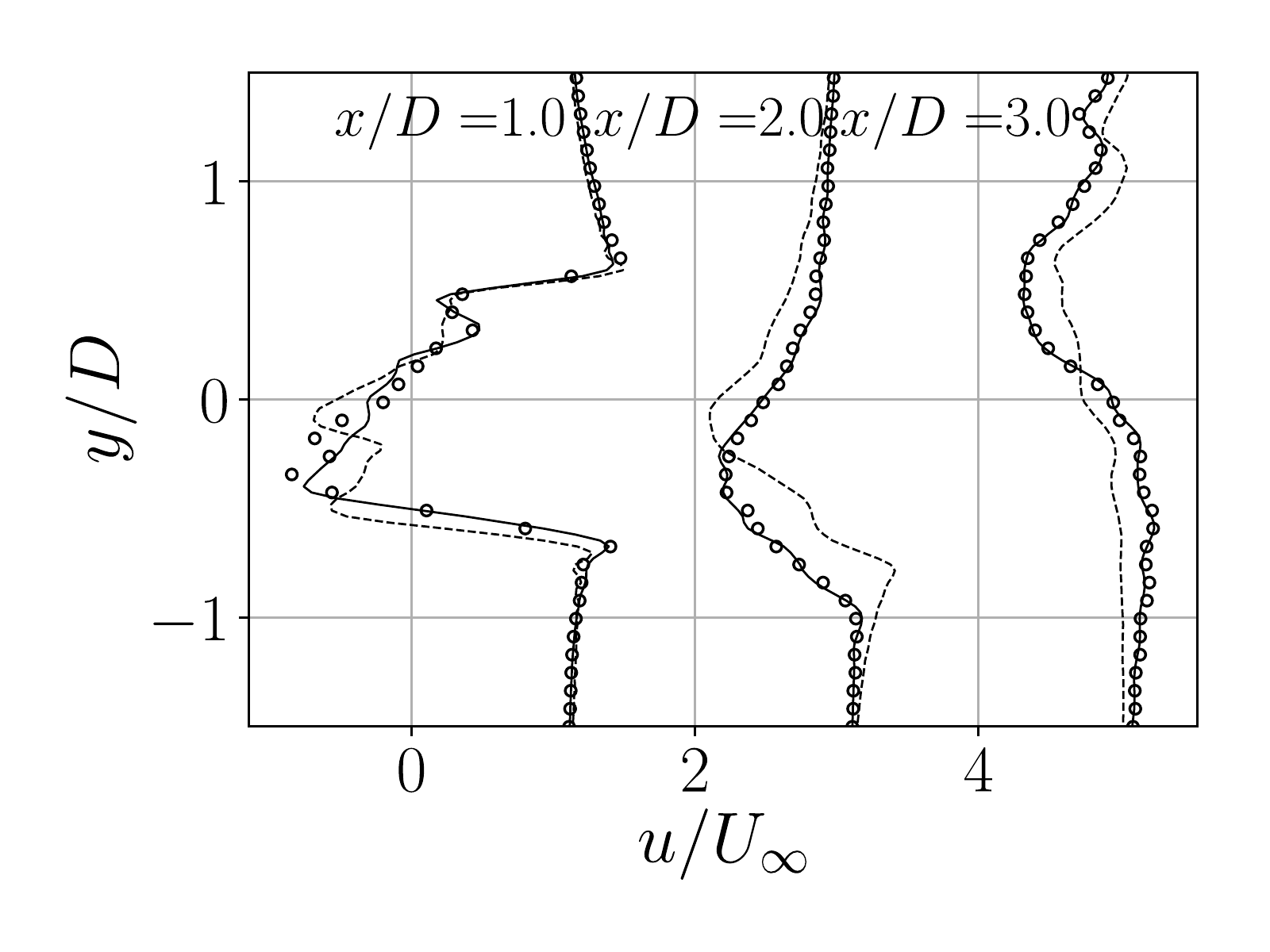}}
  \caption{Profiles of the instantaneous cross-stream velocity along the $x$ axis at $y/D=0$ and $z/D=1.5$ at (a) $Re_{D}=400$ and (b) $Re_{D}=3900$, and the streamwise velocity along the $y$ axis at $x/D=1.0,2.0,3.0$ and $z/D=1.5$ at (c) $Re_{D}=400$ and (d) $Re_{D}=3900$. Dotted lines, from input flow fields at $0\delta t$; solid lines, from flow fields predicted by the CNN system at $5 \delta t$; and circles, from ground truth flow fields at $5 \delta t$. Profiles of the streamwise velocity at $x/D=2.0$ and $3.0$ are shifted by $2.0$ and $4.0$ in the horizontal axis, respectively.}
  \label{fig:line-3900}
\end{figure}

\begin{figure}
  \centering
  \includegraphics[width=0.60\linewidth,trim={0.0cm 0.0cm 0.0cm 0.0cm},clip]{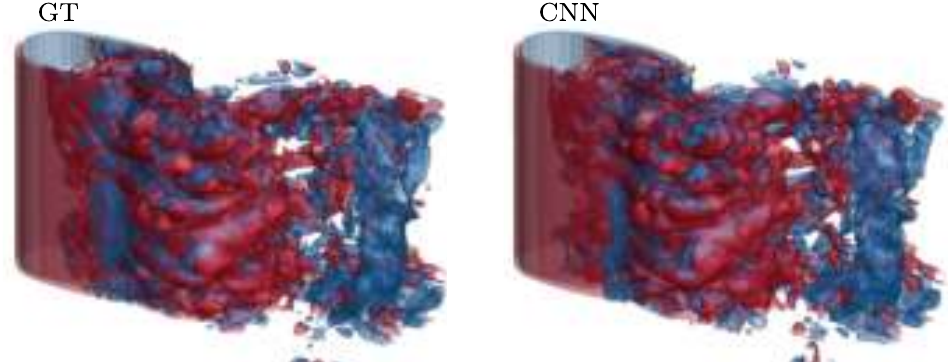}
  \caption{Isosurfaces of the instantaneous streamwise ($\omega_{x}D/U_{\infty}$) and spanwise ($\omega_{z}D/U_{\infty}$) vortices, in the wake of cylinder flow at $5 \delta t$ at $Re_{D}=3900$, calculated from the result predicted by the CNN system (CNN) and from the corresponding ground truth flow field (GT). Red-colored isosurfaces, $\omega_{x}D/U_{\infty}=\omega_{z}D/U_{\infty} = 2.0$; blue-colored isosurfaces, $\omega_{x}D/U_{\infty}=\omega_{z}D/U_{\infty} = -2.0$.}
 \label{fig:iso-vor-3900}
\end{figure}

Wake profiles of the streamwise and cross-stream velocity components calculated from the ground truth flow fields and the predicted flow fields at Reynolds numbers of $Re_{D}=400$ and $3900$ are compared in Fig.~\ref{fig:line-3900}.
As expected, flow at $Re_{D}=3900$, in the shear-layer transition regime, shows larger discrepancy in the wake profiles compared to flow at $Re_{D}=400$ at locations $x/D<2.5$ due to more abundant occurrence of vortex stretching and tilting. However, wake profiles are overall reasonably well predicted by the CNN system.
Especially, convection of flow in the streamwise direction observed from the shift of profiles of the cross-stream velocity (Figs.~\ref{fig:line-3900}(a)~and~(b)) and diffusion of flow observed from the flattening of profiles of the streamwise velocity at downstream locations (Figs.~\ref{fig:line-3900}(c)~and~(d)) are well predicted.
It is also found that small-scale flow structures at $Re_{D}=3900$ are well generated by the CNN system (see Fig.~\ref{fig:iso-vor-3900}).
These results indicate that the same convolution kernels in the CNN system are able to approximate flow dynamics in different flow regimes with various length scales of flow structures.
The capability of the CNN system for predicting flow structures with various length scales implies that a CNN is capable of transporting wavenumber information of flow structures through the network. 

\begin{figure}
  \centering
  \subfigure[]{\includegraphics[width=0.45\linewidth,trim={0.60cm 0.60cm 0.60cm 0.60cm},clip]{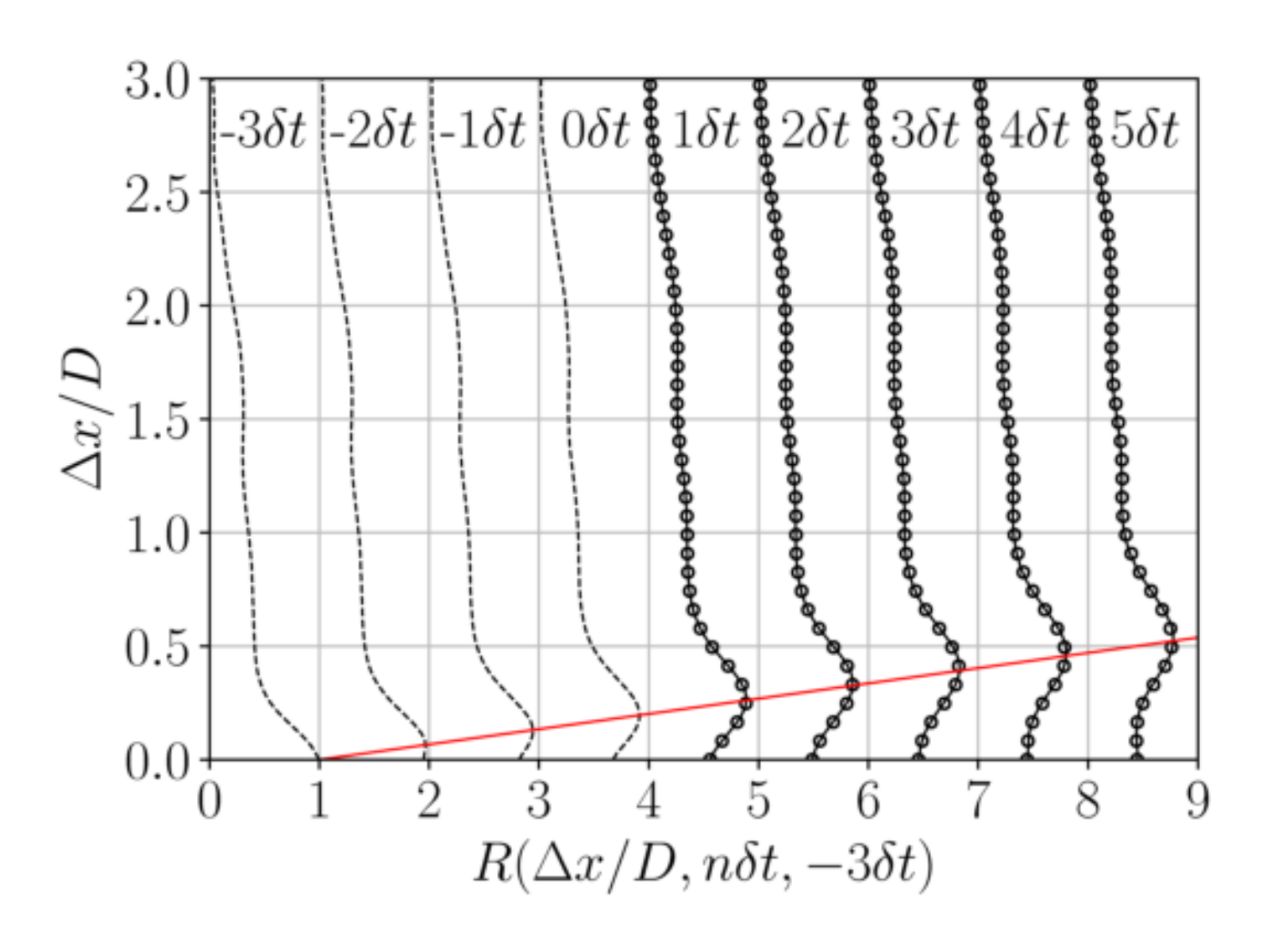}}
  \subfigure[]{\includegraphics[width=0.45\linewidth,trim={0.60cm 0.60cm 0.60cm 0.60cm},clip]{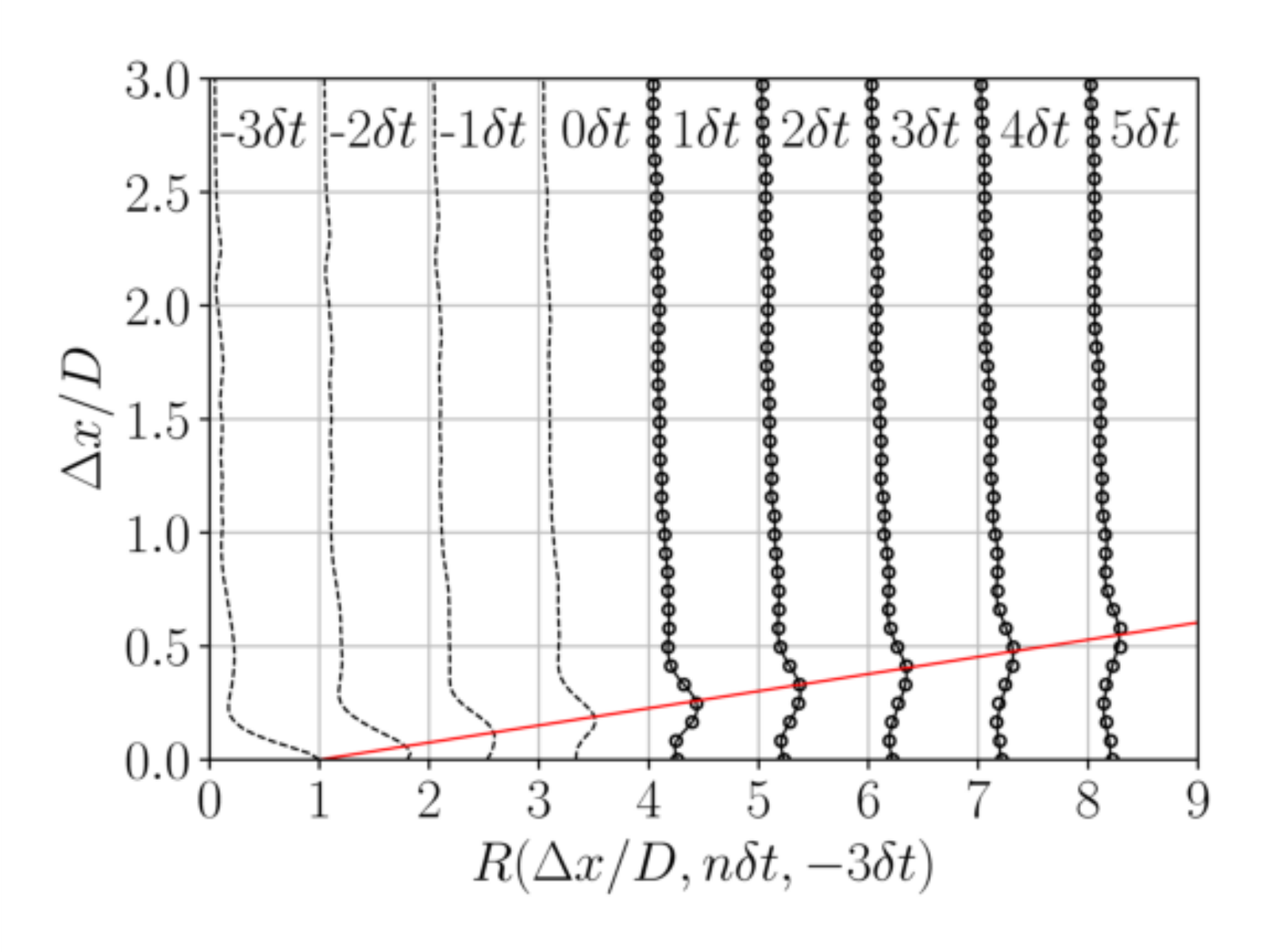}}

  \subfigure[]{\includegraphics[width=0.45\linewidth,trim={0.60cm 0.60cm 0.60cm 0.60cm},clip]{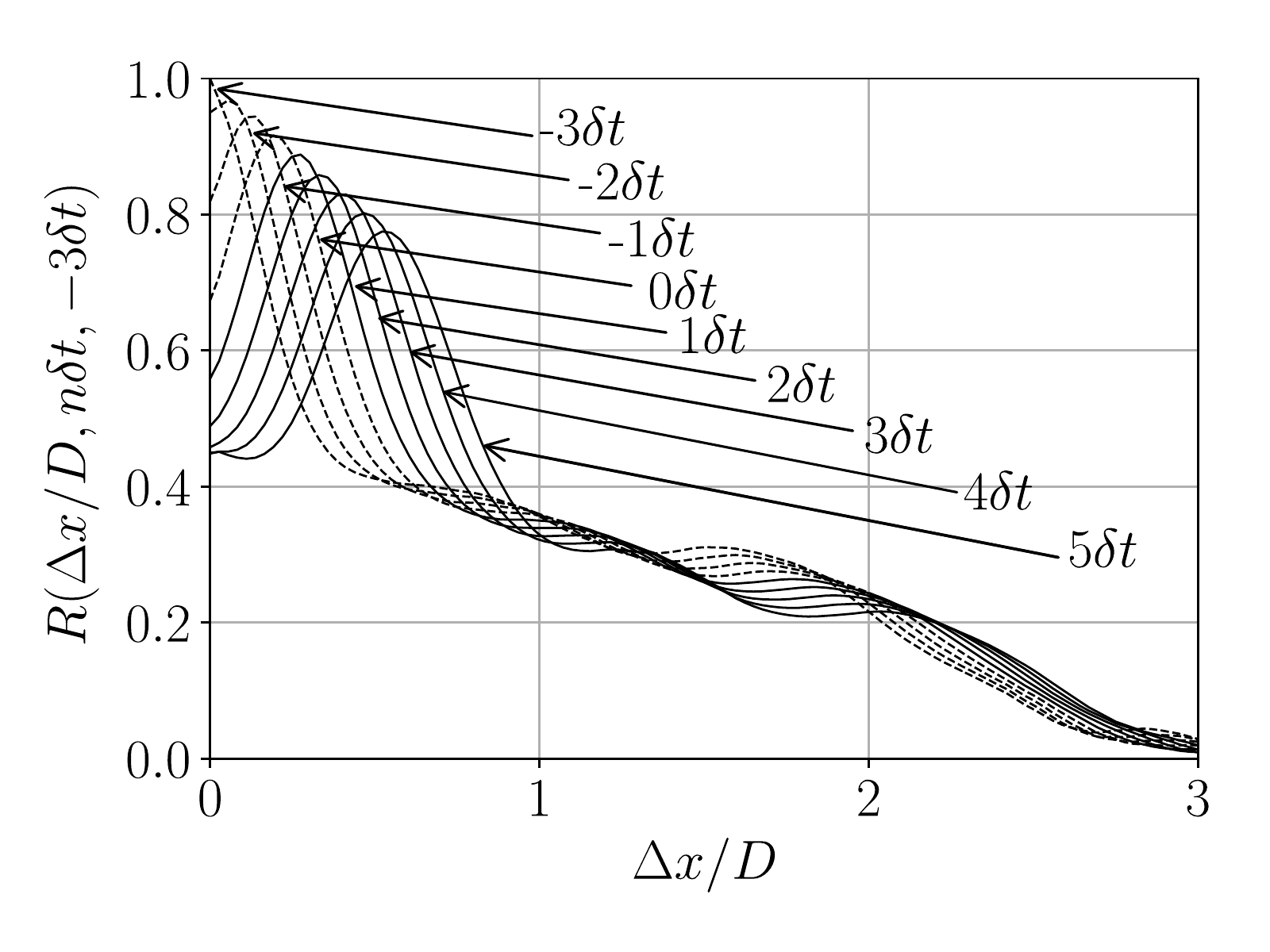}}
  \subfigure[]{\includegraphics[width=0.45\linewidth,trim={0.60cm 0.60cm 0.60cm 0.60cm},clip]{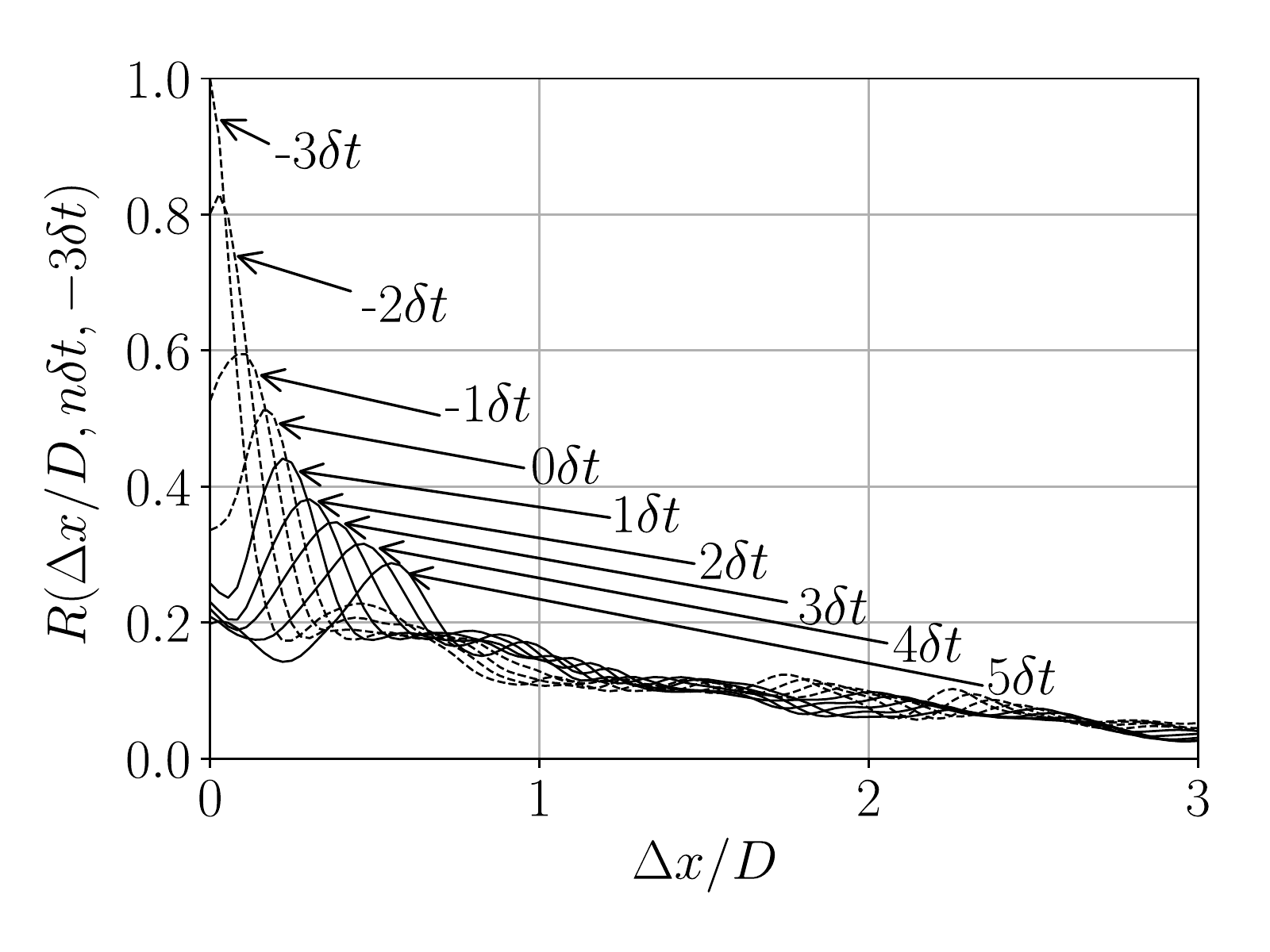}}
  \caption{Space-time correlations of the cross-stream vorticity $R(\Delta x/D, n\delta t, -3\delta t)$ at (a) $Re_{D}=400$ and (b) $Re_{D}=3900$. The space-time correlation at the time step $n \delta t$ is shifted by $n+3$ in the horizontal axis, where $n$ is an integer in [$-3$,$5$]. Red solid lines connect points of the maximum correlations at time steps of $-3\delta t$ and $5\delta t$. Overlapped space-time correlations of the cross-stream vorticity at (c) $Re_{D}=400$ and (d) $Re_{D}=3900$.
  Dotted lines, solid lines, and circles indicate correlations calculated from input flow fields, from flow fields predicted by the  CNN system, and from the ground truth flow fields, respectively.
}
  \label{fig:space-time}
\end{figure}
To investigate temporal dynamics of vortices, space-time correlations of the cross-stream vorticity on the symmetry plane  ($x$-$z$ plane at $y/D=0$) are calculated as follows:
\begin{equation}
R(\Delta x/D, n\delta t, t_{0}) = \frac{<\omega_{y}(x,z,t_{0}) \cdot \omega_{y}(x+\Delta x/D,z,n\delta t)>}{<\omega_{y}(x,z,t_{0}) \cdot \omega_{y}(x,z,t_{0})>},
\end{equation}
where $\omega_{y}(x,z,t_{0})$ is a reference vorticity field at the time step $t_{0}$ and $\omega_{y}(x+\Delta x/D, z, n\delta t)$ is a vorticity field shifted by $\Delta x/D$ in the streamwise direction at the time step of $n\delta t$.
Correlations $R(\Delta x/D, n\delta t, -3\delta t)$ for $\{n\in Z \ | \ -3 \leq n \leq 5\}$ at input time steps ($-3\delta t$ to $0\delta t$) and prediction time steps ($1\delta t$ to $5\delta t$) of flow at $Re_{D}=400$ and $3900$ are shown in Fig.~\ref{fig:space-time}.
The correlation $R(\Delta x/D, n\delta t, -3\delta t)$ at the time step of $n\delta t$ is maximized when the shifted distance $\Delta x/D$ is equal to the average distance of vortices convected to downstream, i.e., convection distance $x^{n\delta t}_{c}$, during time steps from $-3\delta t$ to $n\delta t$.
The convection distance $x^{n\delta t}_{c}$, which corresponds to the maximum correlation $R(x^{n\delta t}_{c}, n\delta t, -3\delta t)$ at the time step of $n\delta t$, is found to increase approximately $0.06D$ per time-step interval $\delta t$ in all input time steps ($-3\delta t$ to $0\delta t$) and prediction time steps ($1\delta t $ to $5\delta t$) (see the red lines in Figs.~\ref{fig:space-time}(a)~and~(b)).
It is worth noting that the convection distance $x^{\delta t}_{c}\sim 0.06D$ per time-step interval $\delta t$ is also identifiable in Fig.~\ref{fig:vor-sym-400}.
The present CNN system is found to be able to predict the convection of vortices by utilizing the temporal information of the convection in input flow fields.

Effects of deformation of vortices due to diffusion, stretching, and tilting during time steps from $-3\delta t$ to $n\delta t$ are inferred in the value of the maximum correlation $R(x^{n\delta t}_{c}, n\delta t, -3\delta t)$ since the maximum correlation at the time step $n\delta t$ is obtained by shifting vortices by the convection distance $x^{n\delta t}_{c}$ in the streamwise direction.
The value of $R(x^{n\delta t}_{c}, n\delta t, -3\delta t)$ at $Re_{D}=400$ is found to decrease linearly with each time-step interval, while the value of $R(x^{n\delta t}_{c}, n\delta t, -3\delta t)$ at $Re_{D}=3900$ is found to decrease faster compared to the decrement observed in the case of $Re_{D}=400$ (see Figs.~\ref{fig:space-time}(c)~and~(d)). The faster decrement of $R(x^{n\delta t}_{c}, n\delta t, -3\delta t)$ at $Re_{D}=3900$ might be partly due to more abundant occurrence of vortex stretching and tilting compared to those in $Re_{D}=400$.
Despite the different decrement characteristics of $R(x^{n\delta t}_{c}, n\delta t, -3\delta t)$ in the input time steps ($-3\delta t$ to $0\delta t$) at Reynolds numbers of $Re_{D}=400$ and $3900$, it is found that the CNN system predicts the trends of the decrements of $R(x^{n\delta t}_{c}, n\delta t, -3\delta t)$ in the prediction time steps ($1\delta t$ to $5\delta t$) at both Reynolds numbers.
The present space-time correlations are considered to imply that the present CNN system is able to predict vortex dynamics, of which characteristics vary in flow at different Reynolds numbers, by extracting temporal information of dynamics such as convection and diffusion phenomena in input flow fields.

\subsubsection{Visualization of information in feature maps}
Feature maps are containers of information of input data or results of stacked convolution and non-linear operations on the input data.
Feature maps in the generative CNN $G_{0}$ of the CNN system are analyzed to discover common structures that the CNN system extracts from the wake of a circular cylinder at Reynolds numbers of $Re_{D}=400$ and $3900$.
Firstly, sizes of information in feature maps in the CNN system are evaluated at both Reynolds numbers.
The size of information $I$ in a three-dimensional feature map $F$ is defined as follows:
\begin{eqnarray}
I = \sqrt{\sum^{n_{x},n_{y},n_{z}}_{i,j,k} F^{2}_{i,j,k}},
\label{eqn:information}
\end{eqnarray}
where, subscripts $i$, $j$, and $k$ are the cell indices and $n_{x}$, $n_{y}$, and $n_{z}$ are numbers of cells in each direction. Then, a relative size of information in a feature map on a layer is evaluated as follows:
\begin{eqnarray}
  I_{rel} = I/I_{max},
\end{eqnarray}
where $I_{max}$ is the maximum size of information in the layer of feature maps.
\begin{figure}
  \centering
  \includegraphics[width=0.7\linewidth,trim={0.0cm 0.0cm 0.0cm 0.0cm},clip]{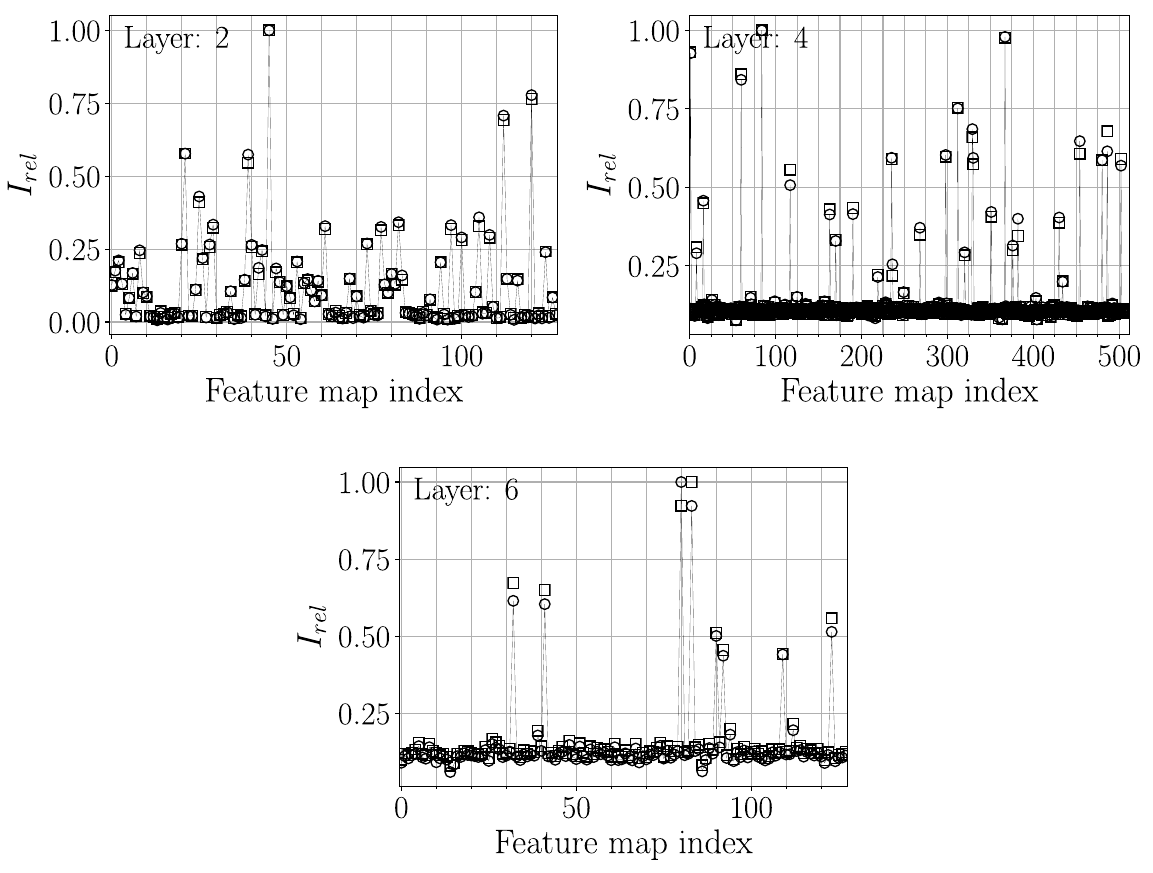}
  \caption{Relative sizes of information in feature maps in the second, fourth, and sixth layers in $G_{0}$. Circles and squares indicate results from Reynolds numbers of $400$ and $3900$, respectively.}
  \label{fig:heatmap}
\end{figure}
The calculated relative sizes of information in feature maps from flow at Reynolds numbers $400$ and $3900$ are shown in Fig.~\ref{fig:heatmap}.
Even though flow fields at Reynolds numbers $400$ and $3900$ are different, magnitudes of the relative size of information in feature maps calculated from both Reynolds numbers are nearly identical.
The similarity of relative sizes of information implies that a feature map which can be considered as an output of composite functions of input flow fields, extracts common flow structures from input flow fields for both Reynolds numbers.

\begin{figure*}
  \centering
  \subfigure[]{\includegraphics[width=0.90\linewidth,trim={0.0cm 0.0cm 0.0cm 0.0cm},clip]{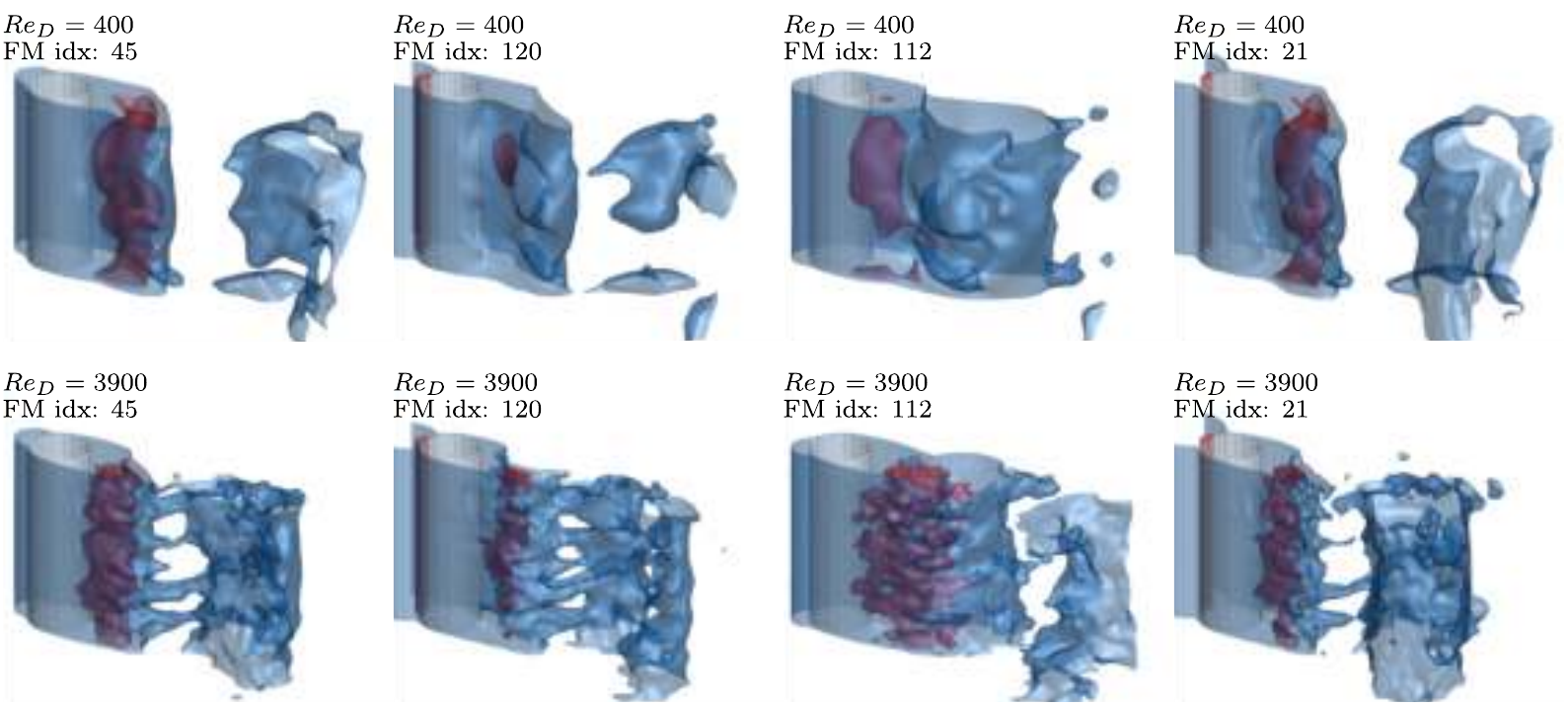}}
  \subfigure[]{\includegraphics[width=0.90\linewidth,trim={0.0cm 0.0cm 0.0cm 0.0cm},clip]{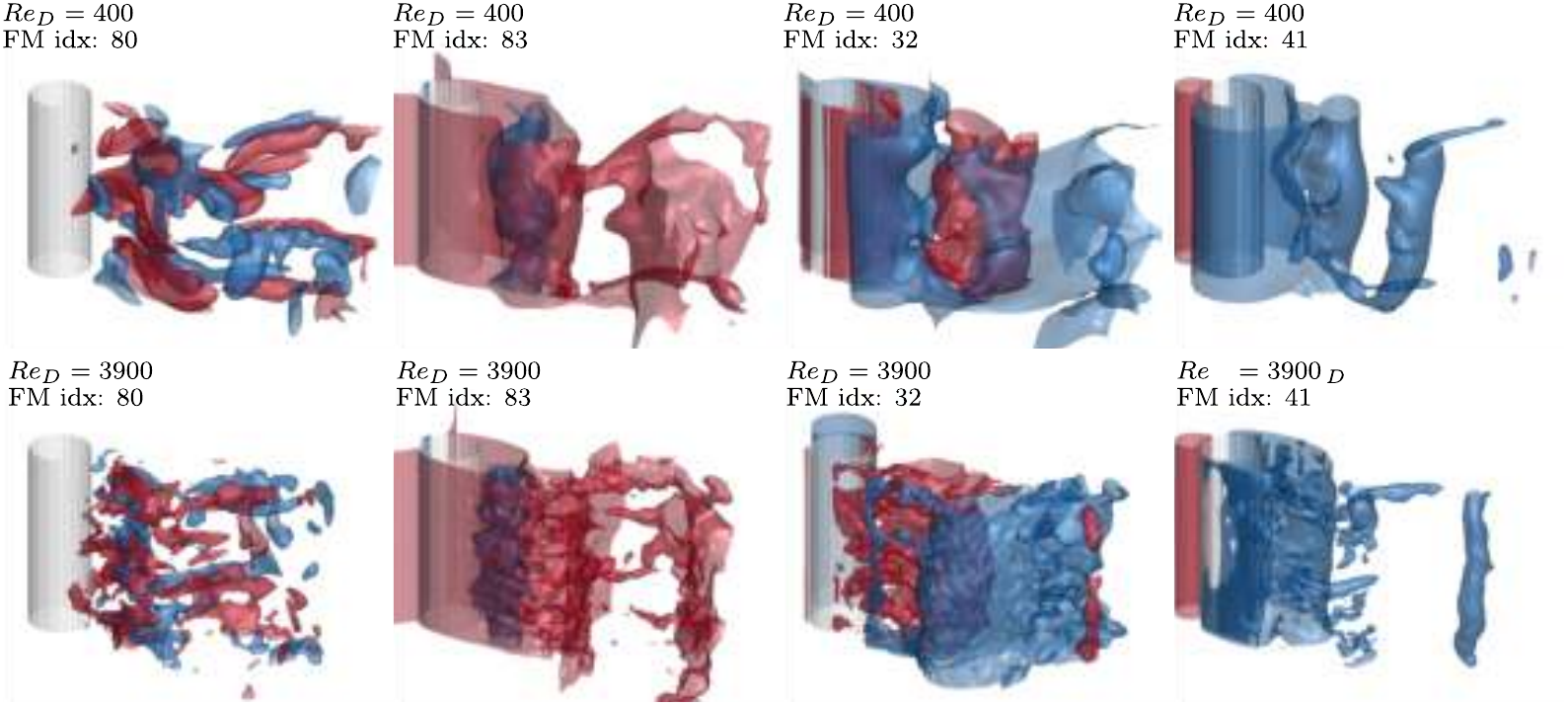}}
  \caption{Flow structures in feature maps extracted from flow at $Re_{D}=400$ and $3900$. Feature maps with the top-four largest relative sizes of information in the (a) second and (b) sixth layers of feature maps in $G_{0}$ are shown. The structures are visualized with isosurface values of 0.3 times of the maximum positive value (red) and the minimum negative value (blue) inside the feature map.}
  \label{fig:FM}
\end{figure*}

Flow structures in feature maps with the top-four largest relative sizes of information at both Reynolds numbers are visualized and compared in Fig.~\ref{fig:FM}. Flow structures in feature maps are visualized with isosurface values relative to the maximum positive value and the minimum negative value inside each feature map, since all feature maps carry information with different ranges of values.
As expected, extracted flow structures at both Reynolds numbers in feature maps with the same indices (FM idx) show similar characteristics. For example, feature maps on the second layer, which are results from the first convolution operations on input flow fields, show sets of structures which resemble braid shear layers and shedding vortices, which are mostly dominated by the streamwise and cross-stream velocity components (Fig.~\ref{fig:FM}(a)). However, feature maps calculated from flow at $Re_{D}=3900$ are found to contain smaller flow structures compared to those at $Re_{D}=400$. For instance, structures related to the streamwise vortices are more observed. The present finding indicates that wavenumber information in input flow fields is transported through convolution layers, and this transportation enables the CNN system to predict flow with length scales that were not utilized during training. Also, feature maps on a deep layer (Fig.~\ref{fig:FM}(b)), which are outputs of composite functions of many convolution operations, show clearer types of flow structures, compared to those on the second layer. 
Composite functions, due to the stacked convolution layers, are expected to integrate and transport flow structures in input flow fields to refine the representation of wake flow dynamics.

\subsubsection{Fourier analysis of convolution kernels}
Mechanisms of a convolution layer to integrate and transport wavenumber information are investigated through a Fourier analysis.
The present CNN system is composed of convolution kernels with the size of $3\times 3 \times 3$.
Let's consider a convolution kernel $\mathcal{W}(x,y,z)$ that connects an input feature map $F(x,y,z)$ and an output feature map $\widetilde{F}(x,y,z)$.
Then, information in the feature map $F(x,y,z)$ is transported to the feature map $\widetilde{F}(x,y,z)$ by a convolution operation ($*$) in space as follows:
\begin{equation}
    \widetilde{F}(x,y,z) = \mathcal{W}(x,y,z) * F(x,y,z).
\label{eqn:conv}
\end{equation}
The convolution kernel $\mathcal{W}(x,y,z)$ comprises kernels associated with convolution operations in the streamwise, cross-stream, and spanwise directions.
Let $\mathcal{W}_{i,j,k}(x,y,z)$ for $i,j,k\in\{1,2,3\}$ be entries of the convolution kernel $\mathcal{W}(x,y,z)$ and $\mathcal{W}_{\_,j,k}(x)$, $\mathcal{W}_{i,\_,k}(y)$, and $\mathcal{W}_{i,j,\_}(z)$ be kernels, which are vectors with the size of $3$, associated with convolution operations in the streamwise, cross-stream, and spanwise directions, respectively.
Note that nine ($=3\times 3$) kernels with the size of $3$ are utilized for convolution operations in each direction.
Now, let's consider a sinc function of $p\in\{x,y,z\}$ with the maximum wavenumber of $k_{0}$ as follows:
\begin{equation}
    sinc(k_{0} p) = \frac{sin(k_{0} p)}{k_{0} p}.
\label{eqn:sinc}
\end{equation}
A Fourier transform of the sinc function Eq.~(\ref{eqn:sinc}) leads to
\begin{equation}
    \widehat{sinc}(k_{p}) = \frac{\pi}{k_{0}} H(|k_{0}|-|k_{p}|),
\label{eqn:sinc-FT}
\end{equation}
where the hat operator $\widehat{()}$ indicates a Fourier transform and $H$ is the Heaviside step function.
By substituting the feature map term of $F(x,y,z)$ in Eq.~(\ref{eqn:conv}) with sinc functions in the streamwise ($x$), cross-stream ($y$), and spanwise ($z$) directions, convolution operations performed by kernels in each direction ($\mathcal{W}_{\_,j,k}(x)$, $\mathcal{W}_{i,\_,k}(y)$, and $\mathcal{W}_{i,j,\_}(z)$) can be calculated as follows:
\begin{eqnarray}
    \widetilde{F}(x) = \sum^{3}_{j=1}\sum^{3}_{k=1} \mathcal{W}_{\_,j,k}(x) * sinc(k_{0x}x), \nonumber \\
    \widetilde{F}(y) = \sum^{3}_{i=1}\sum^{3}_{k=1} \mathcal{W}_{i,\_,k}(y) * sinc(k_{0y}y), \nonumber \\
    \widetilde{F}(z) = \sum^{3}_{i=1}\sum^{3}_{j=1} \mathcal{W}_{i,j,\_}(z) * sinc(k_{0z}z),
\label{eqn:conv-dir}
\end{eqnarray}
where $k_{0x}$, $k_{0y}$, and $k_{0z}$ are the maximum wavenumbers in the streamwise, cross-stream, and spanwise directions, respectively.
The maximum wavenumbers are determined by the Nyquist limit of grid spacings as $k_{0x}=\frac{\pi}{dx}$, $k_{0y}=\frac{\pi}{dy}$, and $k_{0z}=\frac{\pi}{dz}$, where $dx$, $dy$, and $dz$ are grid spacings in each direction.
According to the convolution theorem, Fourier transforms of the convolution operations in Eq.~(\ref{eqn:conv-dir}) are calculated as follows:
\begin{eqnarray}
    \widehat{\widetilde{F}}(k_{x}) = \frac{\pi}{k_{0x}}\sum^{3}_{j=1}\sum^{3}_{k=1} \widehat{\mathcal{W}}_{\_,j,k}(k_{x}) \cdot H(|k_{0x}|-|k_{x}|), \nonumber \\
    \widehat{\widetilde{F}}(k_{y}) = \frac{\pi}{k_{0y}}\sum^{3}_{i=1}\sum^{3}_{k=1} \widehat{\mathcal{W}}_{i,\_,k}(k_{y}) \cdot H(|k_{0y}|-|k_{y}|), \nonumber \\
    \widehat{\widetilde{F}}(k_{z}) = \frac{\pi}{k_{0z}}\sum^{3}_{i=1}\sum^{3}_{j=1} \widehat{\mathcal{W}}_{i,j,\_}(k_{z}) \cdot H(|k_{0z}|-|k_{z}|). \nonumber \\
\label{eqn:conv-dir-FT}
\end{eqnarray}
As the Heaviside step functions are unity for wavenumbers smaller than the Nyquist limits, $\widehat{\widetilde{F}}(k_{x})$, $\widehat{\widetilde{F}}(k_{y})$, and $\widehat{\widetilde{F}}(k_{z})$ represent the transported wavenumber information from the convolution kernels in each direction.

\begin{figure}
  \centering
  \subfigure[]{\includegraphics[width=0.35\linewidth,trim={0.5cm 0.5cm 0.5cm 0.5cm},clip]{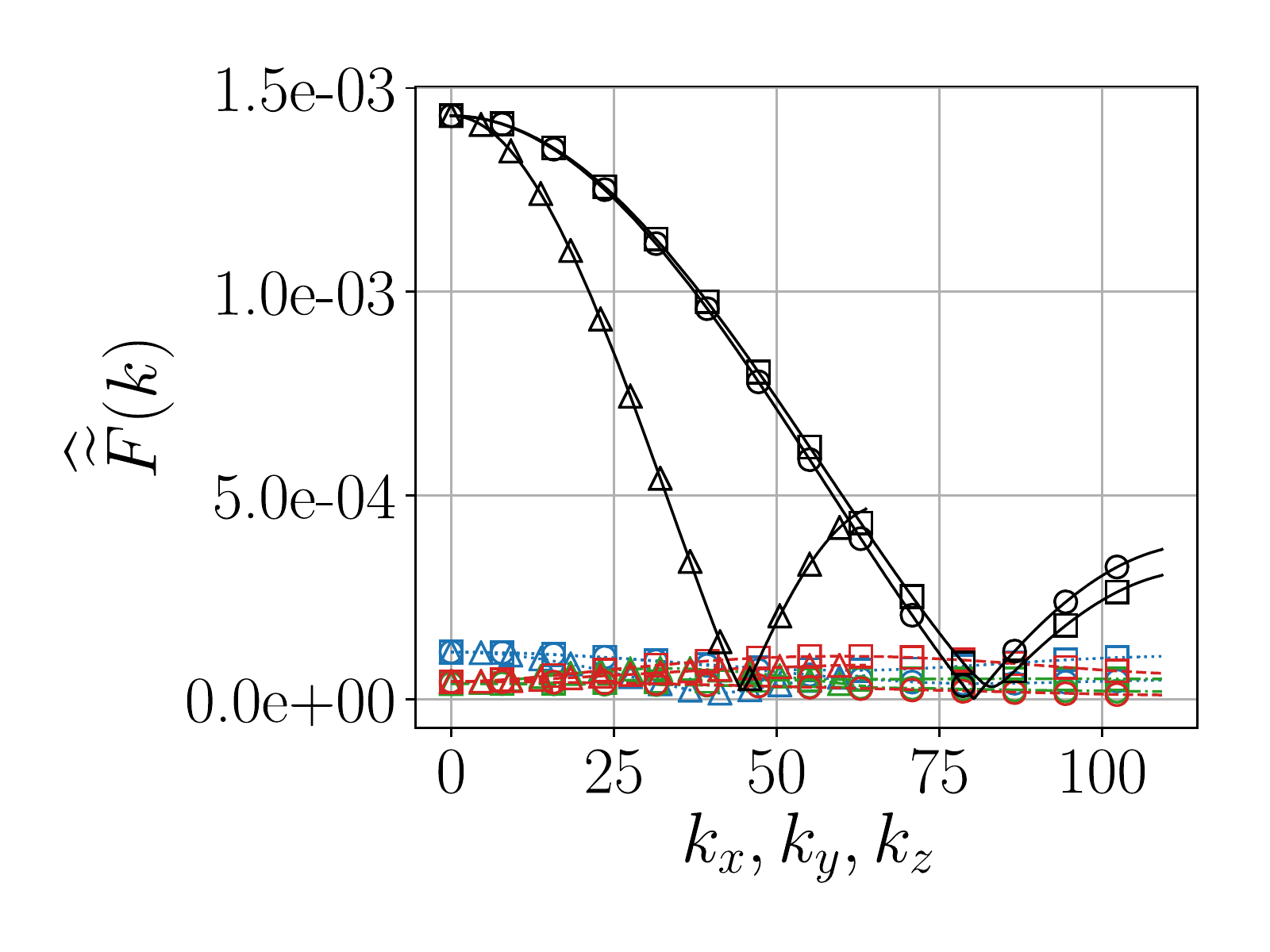}}
  \subfigure[]{\includegraphics[width=0.35\linewidth,trim={0.5cm 0.5cm 0.5cm 0.5cm},clip]{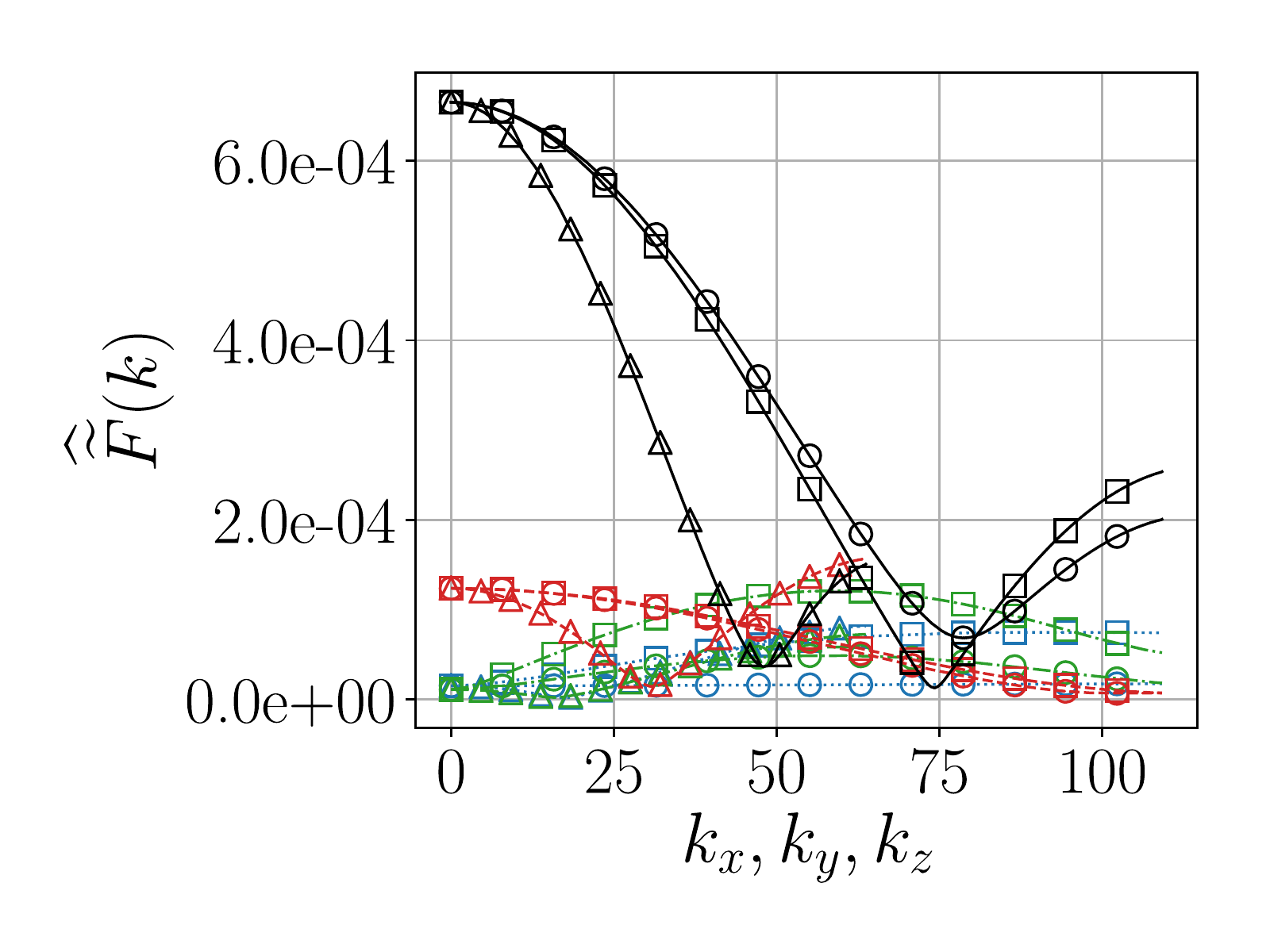}}

  \subfigure[]{\includegraphics[width=0.35\linewidth,trim={0.5cm 0.5cm 0.5cm 0.5cm},clip]{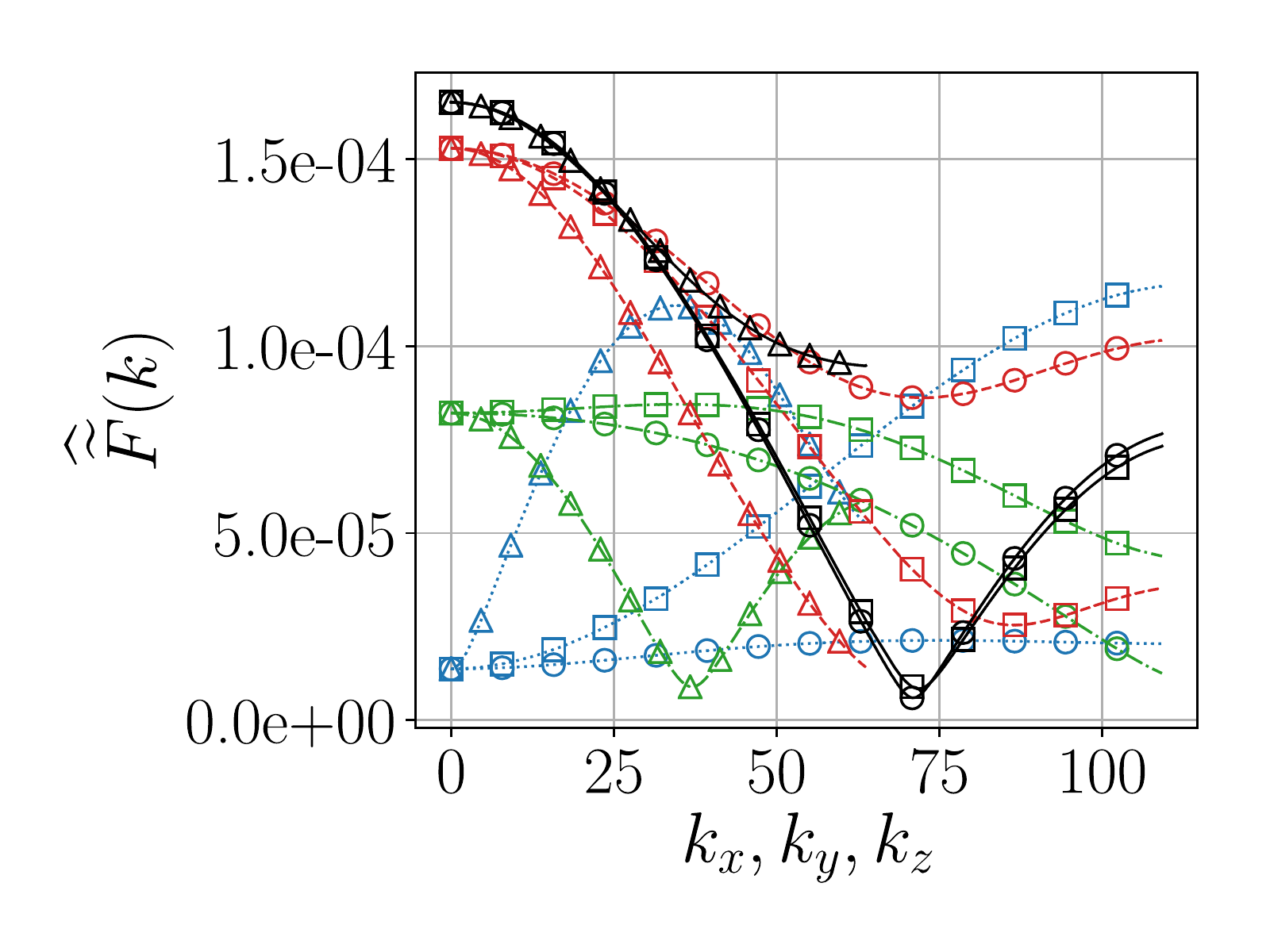}}
  \subfigure[]{\includegraphics[width=0.35\linewidth,trim={0.5cm 0.5cm 0.5cm 0.5cm},clip]{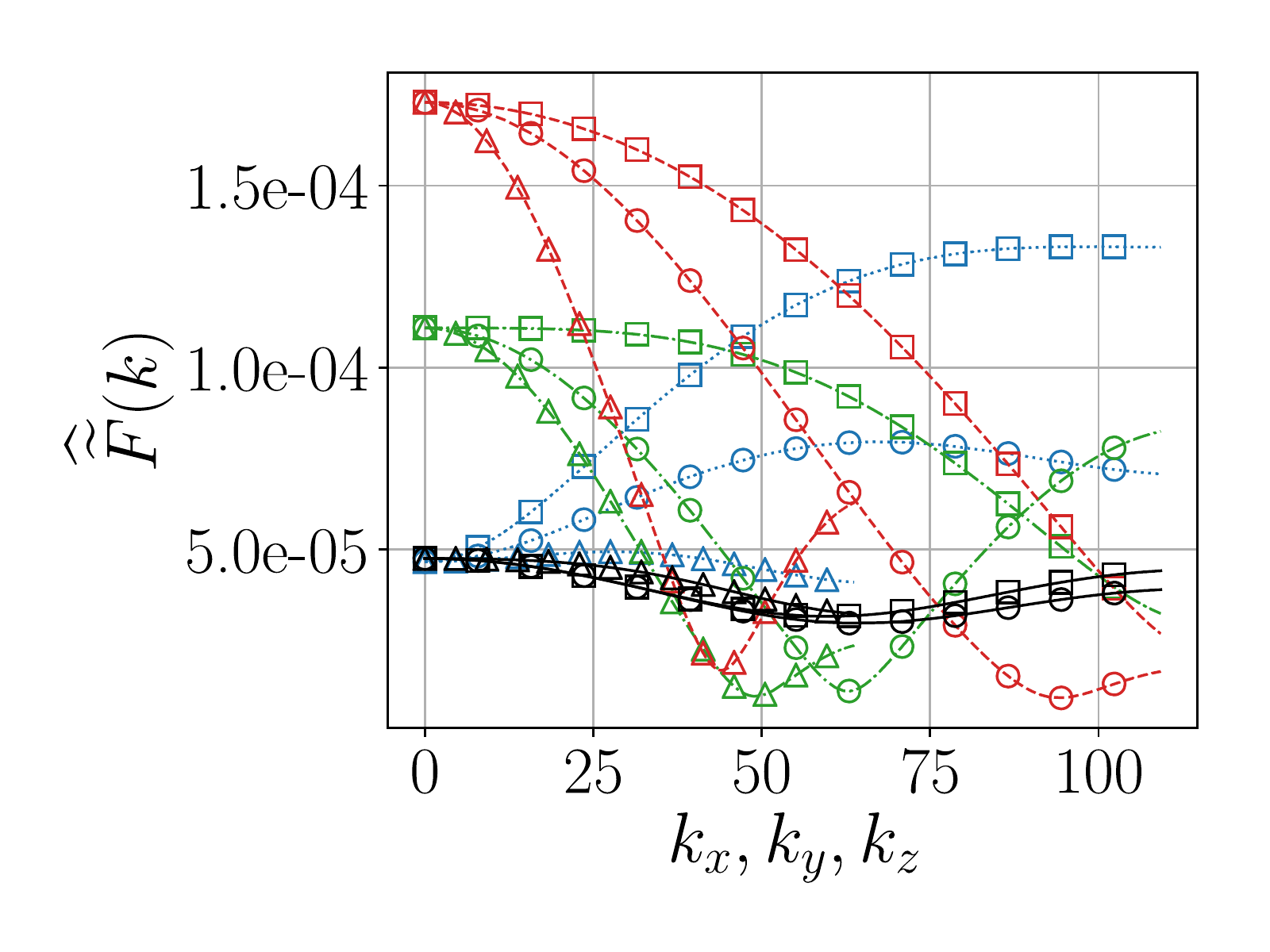}}
  \caption{Transported wavenumber information from convolution kernels that connect the input feature maps of flow variables of (a) $u/U_{\infty}$, (b) $v/U_{\infty}$, (c) $w/U_{\infty}$, and (d) $p/\rho U^{2}_{\infty}$ with the feature map with the largest relative size of information in the second layer of feature maps in $G_{0}$ (see FM idx: 45 in Fig.~\ref{fig:FM}(a) for the flow structures in the feature map). Black solid lines, red dashed lines, green dash-dotted lines, and blue dotted lines indicate the transported wavenumber information from feature maps corresponding to flow histories of $0\delta t$, $-1\delta t$, $-2\delta t$, and $-3\delta t$, respectively. {\Large$\circ$}, $\square$, and $\triangle$ indicate Fourier transforms in the streamwise ($k_{x}$), cross-stream ($k_{y}$), and spanwise ($k_{z}$) directions, respectively.}
  \label{fig:W-top}
\end{figure}
Figure~\ref{fig:W-top} shows the wavenumber information transported by convolution kernels that connect input feature maps of flow variables $\{u$, $v$, $w$, $p\}$ at all flow histories $\{-3\delta t$, $-2\delta t$, $-1\delta t$, $0\delta t\}$ with the feature map with the largest relative size of information in the second layer of feature maps in $G_{0}$.
Convolution kernels are found to function as wavenumber filters, such as low-pass, high-pass, band-pass, and band-stop filters, that cover wavenumbers under the Nyquist limit of the grid.
For instance, convolution kernels connected with input feature maps of the streamwise ($u/U_{\infty}$) and cross-stream ($v/U_{\infty}$) velocity components at the most recent flow history ($0\delta t$) are found to be band-stop type filters which transport large magnitudes of wavenumber components in certain ranges of the wavenumber.
While, other convolution kernels are found to be wavenumber filters that transport comparable magnitudes of wavenumber components in all ranges of the wavenumber.
As a convolution layer sums results of convolution operations on all input feature maps, wavenumber components transported by convolution kernels are integrated into the output feature map.
Integration of wavenumber information from all input feature maps is essential as merged information of flow variables and histories is required to predict flow dynamics.
The process of integration is repeated in the CNN system containing stacked convolution layers, which leads to a further integration and transportation of wavenumber information.
A closer investigation of the transportation characteristics of information of flow variables and histories to deeper layers are discussed in the following sections.

\subsubsection{Transportation characteristics of flow variables and histories}
Transportation characteristics of flow variables are investigated by calculating contributions of flow variables to feature maps in the generative CNN $G_{0}$. Contributions from an input flow variable to feature maps can be calculated by feeding the CNN system with a single input flow variable, while information of the other input flow variables is zeroed out. As zero values cannot travel through a network, this method enables to show pure contributions from each flow variable to feature maps.
Let $I_{u}$, $I_{v}$, $I_{w}$, and $I_{p}$ be information calculated from the CNN system fed with a single input flow variable of $u$, $v$, $w$, and $p$, respectively, using Eq.~(\ref{eqn:information}).
Then, the contribution factor of a flow variable $CF^{var}_{f}$ on a feature map is calculated as follows:
\begin{eqnarray}
CF^{var}_{f} = \frac{I_{f}}{I_{u}+I_{v}+I_{w}+I_{p}},
\label{eqn:con-var}
\end{eqnarray}
where $f \in \{u,v,w,p\}$.
\begin{figure}
  \centering
  \includegraphics[width=0.96\linewidth,trim={0.0cm 0.0cm 0.0cm 0.0cm},clip]{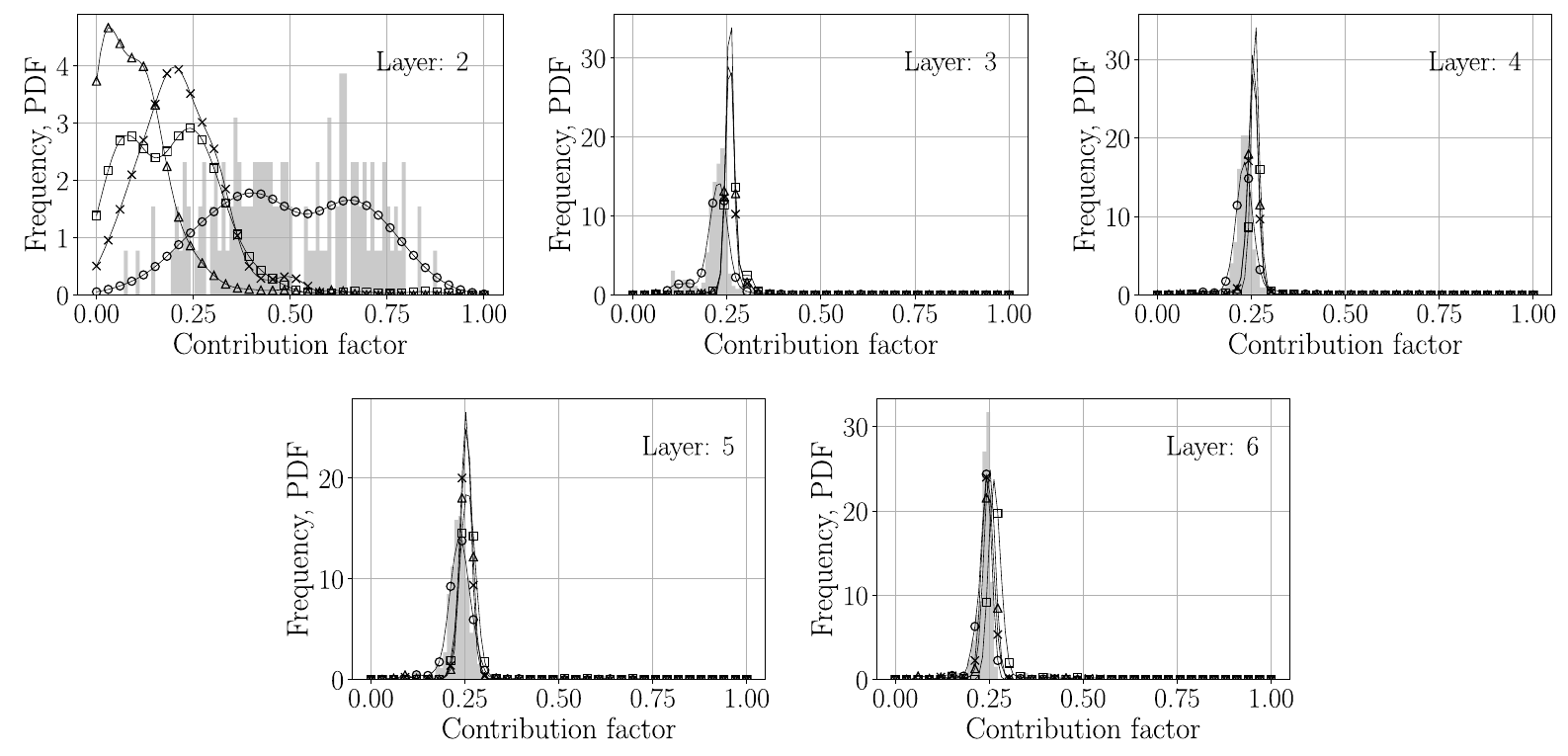}
  \caption{PDFs of input flow variables as a function of contribution factors evaluated on feature maps from the second to the sixth layers in $G_{0}$. {\Large$\circ$}, $CF^{var}_{u}$; $\square$, $CF^{var}_{v}$; $\triangle$, $CF^{var}_{w}$; $\times$, $CF^{var}_{p}$. Histograms of frequencies of $CF^{var}_{u}$ are shown.}
  \label{fig:con-var}
\end{figure}
The calculated frequency distributions and probability density functions (PDFs) of $CF^{var}_{f}$ on feature maps of flow at $Re_{D}=400$ are shown in Fig.~\ref{fig:con-var}.
Information of the streamwise velocity is observed to be the most extracted information of flow variables in the second layer of feature maps since the PDF of $CF^{var}_{u}$ in the second layer remains to be non-zero even within a range, such as $0.50<CF^{var}_{u}<0.75$, where contributions from other flow variables show nearly zero probability densities. However, as the input information of flow variables is transported into deeper layers, PDFs from all flow variables become similar by showing the maximum PDFs on contribution factors near $0.25$, which corresponds to one divided by the number of flow variables.
The present observation indicates that the present CNN system firstly extracts information mostly related to the streamwise velocity, which is the dominant flow variable in dynamics of flow over a circular cylinder, and as the layer goes deeper, the CNN system combines information from all flow variables to extract dynamics related to all flow variables.

Besides to information of flow variables, information of flow histories is essential as it provides temporal features of flow dynamics.
\begin{figure}
  \centering
  \includegraphics[width=0.96\linewidth,trim={0.0cm 0.0cm 0.0cm 0.0cm},clip]{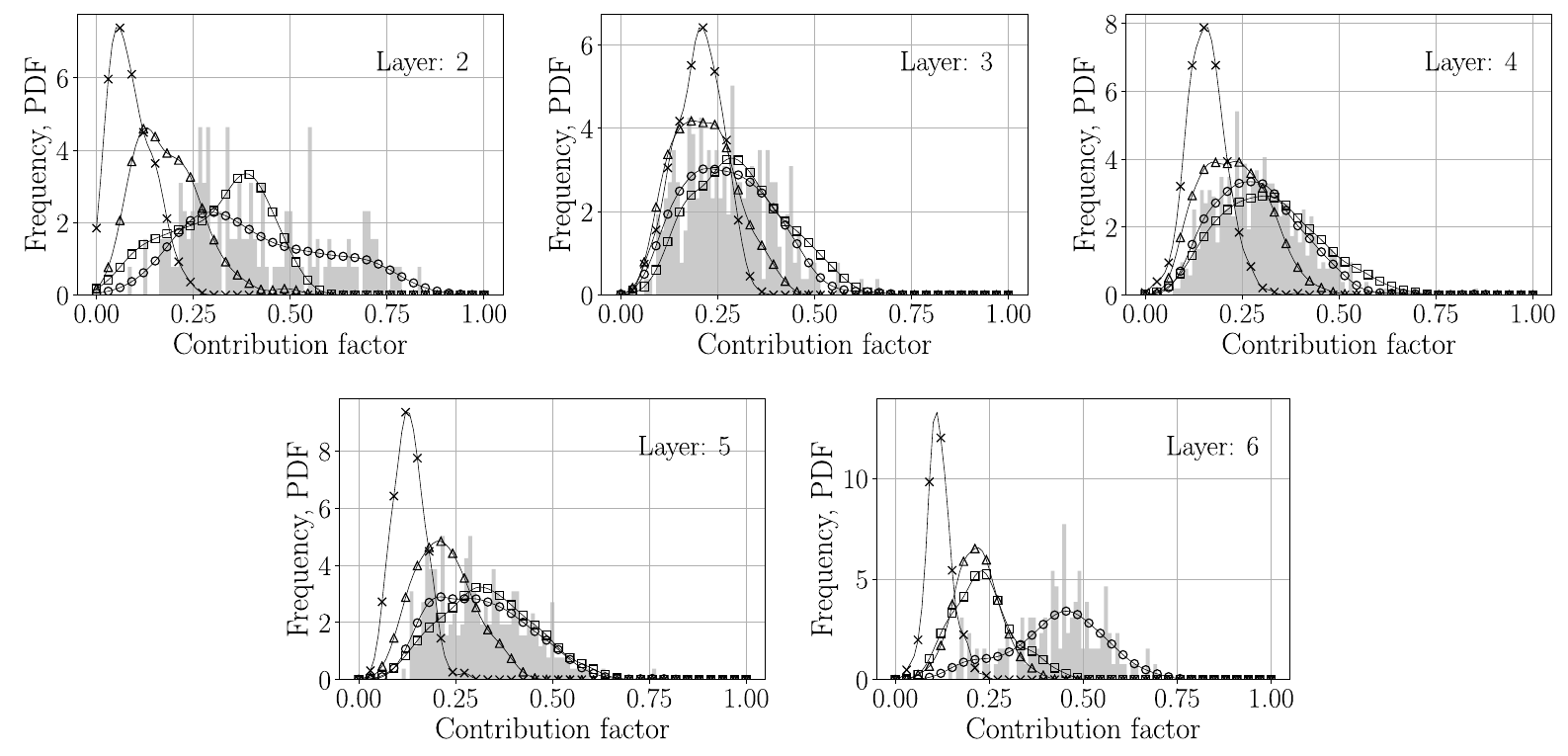}
  \caption{PDFs of input flow histories as a function of contribution factors evaluated on feature maps from the second to the sixth layers in $G_{0}$. {\Large$\circ$}, $CF^{hist}_{0\delta t}$; $\square$, $CF^{hist}_{-1\delta t}$; $\triangle$, $CF^{hist}_{-2\delta t}$; $\times$, $CF^{hist}_{-3 \delta t}$. Histograms of frequencies of $CF^{hist}_{0\delta t}$ are shown.}
  \label{fig:con-hist}
\end{figure}
The transportation characteristics of input flow histories are investigated by calculating the contribution factor of flow histories $CF^{hist}_{f}$ similarly to the contribution factor of flow variables as follows:
\begin{eqnarray}
CF^{hist}_{f} = \frac{I_{f}}{I_{-3 \delta t}+I_{-2 \delta t}+I_{-1 \delta t}+I_{0 \delta t}},
\label{eqn:con-hist}
\end{eqnarray}
where  $f \in \{-3 \delta t,-2 \delta t,-1 \delta t, 0 \delta t\}$ and $I_{-3\delta t}$, $I_{-2\delta t}$, $I_{-1\delta t}$, and $I_{0\delta t}$ are information in a feature map calculated from the CNN system fed with a single set of flow histories at $-3\delta t$, $-2\delta t$, $-1\delta t$, and $0\delta t$, respectively, while information of the other input flow histories is zeroed out. 

Calculated frequency distributions and probability density functions of $CF^{hist}_{f}$ on feature maps of flow at $Re_{D}=400$ are shown in Fig.~\ref{fig:con-hist}.
Sparse contributions from input flow histories, especially from the most recent flow history ($0\delta t$), are observed in the feature maps on the second layer.
However, at feature maps on an intermediate level of layers (Layers 3 to 5 in Fig.~\ref{fig:con-hist}), contributions from input flow histories, especially of $CF^{hist}_{0\delta t}$ and $CF^{hist}_{-1\delta t}$, become similar.
This similarity of contributions implies that feature maps combine information from input flow histories with nearly the same importance, therefore, representation of flow structures with consideration of temporal dynamics are extracted on these feature maps.
As the information is transported into a deeper layer (Layer 6 in Fig.~\ref{fig:con-hist}), contributions of information from the most recent flow history become the largest, while contributions from other flow histories decrease as time distances to the prediction increase.
The characteristics of the contribution factor $CF^{hist}_{f}$ implies that the CNN system learns temporal correlations in unsteady flow dynamics, as temporal correlations of flow fields generally decrease as time distances between flow fields increase.

\subsection{Reduction of the number of feature maps in the CNN system}
Optimizing a CNN or a CNN system which consists of multiple CNNs has always been a difficult task, as the number of feature maps in a CNN has been mostly determined by an extensive parameter study, which demands a high cost for computing many combinations of numbers of feature maps on all layers in a CNN.
In this section, a systematic approach for reducing numbers of feature maps in the CNN system is presented, The present approach is based on the careful observation on feature maps in the CNN system, rather than an extensive parameter study.
\begin{figure}
  \centering
  \subfigure[]{\includegraphics[width=0.60\linewidth,trim={0.0cm 0.0cm 0.0cm 0.0cm},clip]{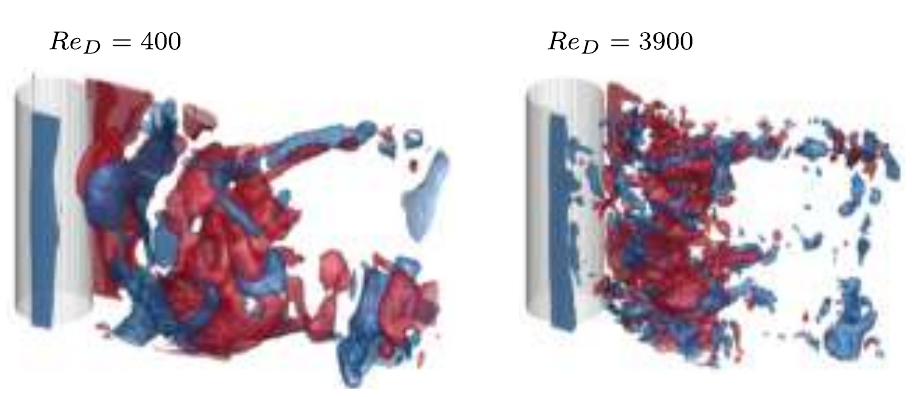}}

  \subfigure[]{\includegraphics[width=0.60\linewidth,trim={0.0cm 0.0cm 0.0cm 0.0cm},clip]{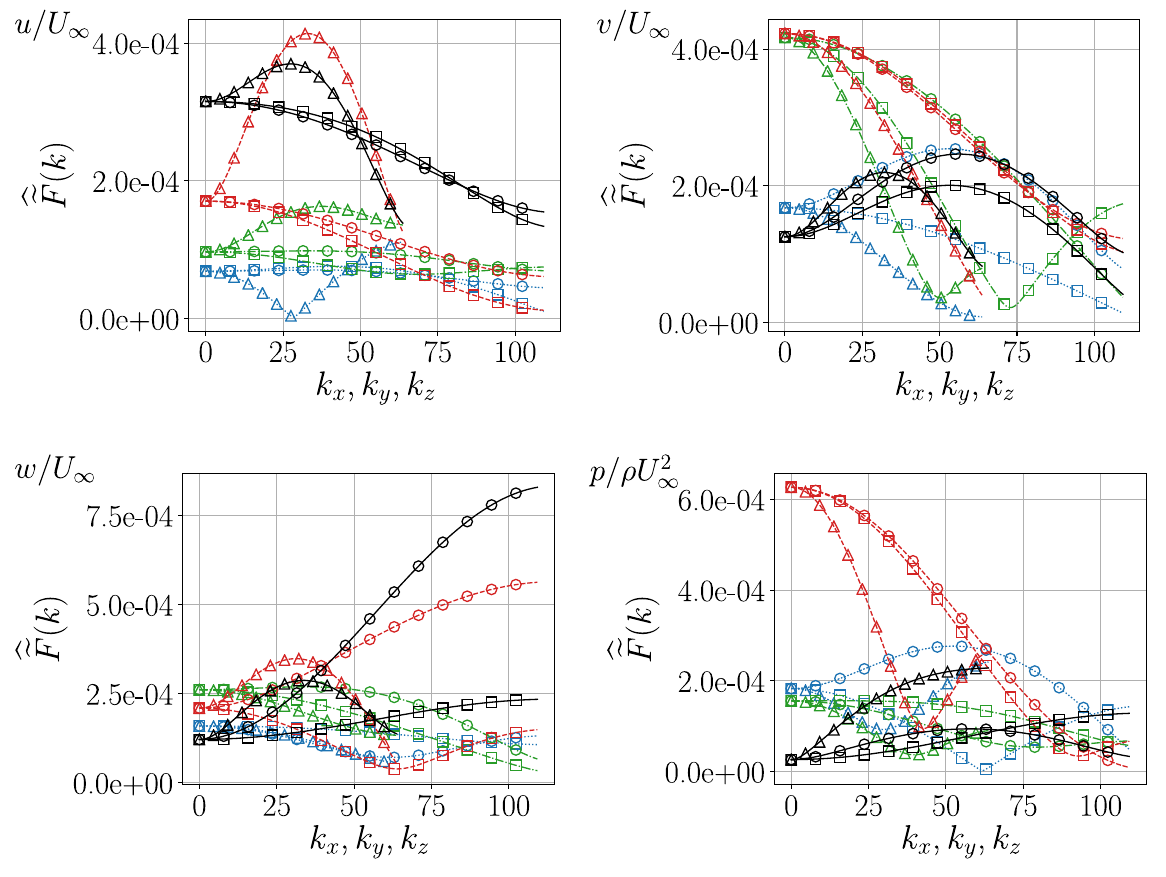}}
  \caption{(a) Flow structures in feature maps, calculated from flow at $Re_{D}=400$ and $3900$, with the third smallest relative size of information in the second layer of feature maps in $G_{0}$ are shown. The structures are visualized with isosurface values of 0.2 times of the maximum positive value (red) and the minimum negative value (blue) inside a feature map. (b) Transported wavenumber information from convolution kernels that connect input feature maps of flow variables of $u/U_{\infty}$, $v/U_{\infty}$, $w/U_{\infty}$, and $p/\rho U^{2}_{\infty}$ with the feature map with the third smallest relative size of information in the second layer of feature maps. Black solid lines, red dashed lines, green dash-dotted lines, and blue dotted lines indicate  transported wavenumber information from feature maps corresponding to flow histories of $0\delta t$, $-1\delta t$, $-2\delta t$, and $-3\delta t$, respectively. {\Large$\circ$}, $\square$, and $\triangle$ indicate Fourier transforms in the streamwise ($k_{x}$), cross-stream ($k_{y}$), and spanwise ($k_{z}$) directions, respectively.}
  \label{fig:FM-small-info}
\end{figure}

Firstly, it is attempted to reduce the number of feature maps based on relative sizes of information in feature maps, since sparse existence of feature maps with relatively large sizes of information and abundant feature maps with relatively small sizes of information are observed (see Fig.~\ref{fig:heatmap}).
However, it is found that even a feature map with a small size of information contains clear flow structures. For instance, the feature map with the third lowest relative size of information on the second layer of feature maps in $G_{0}$, shows structures related to shear layers and wake vortices as shown in Fig.~\ref{fig:FM-small-info}(a).
Also, this feature map is observed to be transported with wavenumber components with comparable magnitudes from all flow variables and histories as shown in Fig.~\ref{fig:FM-small-info}(b).
This observation indicates that the feature map contains information of flow dynamics integrated from all flow variables and histories.
Since even a feature map with a small relative size of information could contain clear flow structures affected by all flow variables and histories, feature maps cannot be simply removed based on the size of information.

\begin{figure}
  \centering
  \subfigure[]{\includegraphics[width=0.30\linewidth,trim={0.0cm 0.0cm 0.0cm 0.0cm},clip]{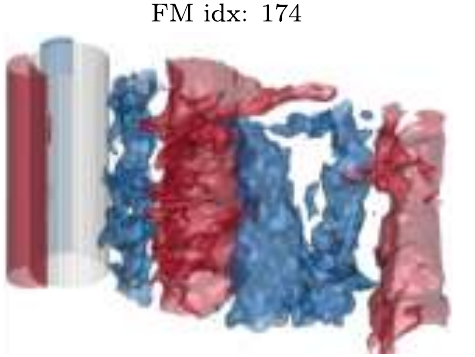}}
  \subfigure[]{\includegraphics[width=0.30\linewidth,trim={0.0cm 0.0cm 0.0cm 0.0cm},clip]{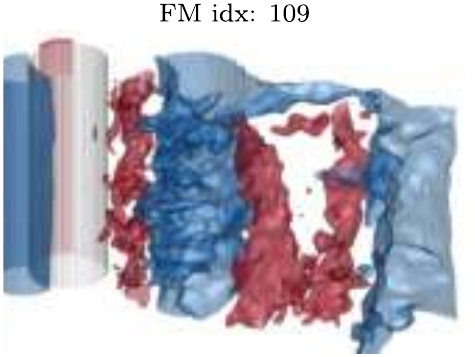}}
  \subfigure[]{\includegraphics[width=0.30\linewidth,trim={0.0cm 0.0cm 0.0cm 0.0cm},clip]{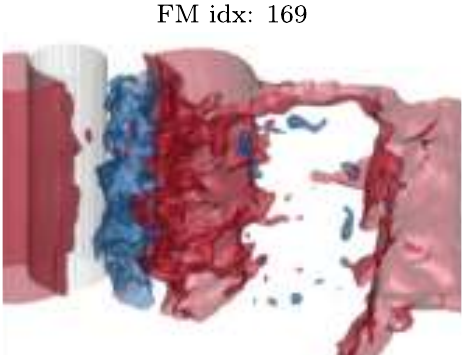}}
  \caption{Flow structures in feature maps with indices of (a) 174, (b) 109, and (c) 169 on the fifth layer in $G_{0}$, extracted from flow at $Re_{D}=3900$. The SSIM value between feature maps of (a) and (b) is $0.8$, while the SSIM value between feature maps of (a) and (c) is $0.7$. Flow structures are visualized with isosurface values of 0.3 times of the maximum positive value (red) and the minimum negative value (blue) inside the feature map.}
  \label{fig:FM-sim}
\end{figure}

However, feature maps that contain redundant flow structures are found in the present CNN system (see Figs.~\ref{fig:FM-sim}(a)~and~(b) for an example).
Therefore, it is attempted to reduce the number of feature maps by evaluating similarities between feature maps.
Similarities between feature maps are quantitatively evaluated using a measure of structural similarity (SSIM) developed by~\citet{wang2004image}.
The SSIM between spatial data of $a$ and $b$ is defined as follows:
\begin{eqnarray*}
SSIM(a,b) = \frac{(2\mu_{a} \mu_{b} +c_{1})(2\sigma_{ab}+c_{2})}{(\mu^{2}_{a}+\mu^{2}_{b}+c_{1})(\sigma^{2}_{a}+\sigma^{2}_{b}+c_{2})}, \nonumber \\
\label{eqn:SSIM-def}
\end{eqnarray*}
where $\mu_{a}$ and $\mu_{b}$ are averages of $a$ and $b$, $\sigma^{2}_{a}$ and $\sigma^{2}_{b}$ are variances of $a$ and $b$, $\sigma_{ab}$ is the covariance of $a$ and $b$, and $c_{1}=(0.01)^{2}$ and $c_{2}=(0.03)^{2}$ are constant values to stabilize a division with weak denominator.
A similarity matrix $S^{m,n}$ of which elements indicate SSIM values between absolutes of feature maps of $F^{m}$ and $F^{n}$ in the same layer, is calculated as follows:
\begin{eqnarray}
S^{m,n} = SSIM(abs(F^{m}),abs(F^{n})).
\label{eqn:SSIM}
\end{eqnarray}
The absolutes of feature maps are utilized in Eq.~(\ref{eqn:SSIM}) to consider similarities between feature maps with opposite signs, since feature maps with opposite signs are possible to transport the same information while the signs are carried by the connected convolution kernels.
The measure of SSIM is expected to well compare flow structures in feature maps, as it is reported to be capable of comparing structural information in data~\cite{wang2004image}.

Here, the process to reduce the number of feature maps in the CNN system based on measures of similarity matrices is explained.
Firstly, similarity matrices in the CNN system are calculated for flow at $Re_{D}=3900$ in the domain of $-1.5<x/D<5.5, -3.5<y/D<3.5$, and $0<z/D<1.5$.
Then, to detect redundant feature maps, proportion of elements $P(T)$ with SSIM values larger than a threshold SSIM value $T$ in a similarity matrix $S^{m,n}$ is calculated as follows:
\begin{equation}
P(T) = \frac{count[(m,n) | S^{m,n} > T]}{\sum_{m,n}},
\label{eqn:Pl}
\end{equation}
where $\sum_{m,n}$ indicates the number of elements in the similarity matrix. A threshold SSIM value of $T=0.8$ is chosen in the present study as feature maps with an SSIM value under $0.8$ tend to contain different flow structures. Figure~\ref{fig:FM-sim} shows an example of feature maps with SSIM values of $0.8$ and $0.7$. Feature maps with an SSIM value of $0.8$ show similar flow structures, while feature maps with an SSIM value of $0.7$ show differences in flow structures. Based on the calculated $P(T)$ in each layer, numbers of feature maps are reduced. For layers with $P(T)$ over $3/4$, which indicates that over three-quarters of feature maps are similar to each other, it is possible to reduce the numbers of feature maps in the layers by three-quarters. Similarly, it is possible to reduce the numbers of feature maps by a half in the layers with $P(T)$ over $1/2$ and under $3/4$. Resulting numbers of feature maps in layers of the CNN system with the reduced number of feature maps, which is referred to the reduced CNN system, are shown in Table~\ref{tab:reduced-CNN}. In the reduced CNN system, the number of parameters to learn, which is proportional to the sum of products of numbers of feature maps from every two adjacent layers in the CNN, is decreased by $85\%$ from those in the original CNN system of which configuration is summarized in Table~\ref{tab:CNN}.
\begin{table}
\centering
\begin{tabular}{c|l}
\hline \hline
\multicolumn{1}{c|}{Generative CNN} & \multicolumn{1}{c}{Numbers of feature maps} \\ \hline
\multicolumn{1}{c|}{$G_{3}$}             & 16, 64, 64, 64, 4\\ 
\multicolumn{1}{c|}{$G_{2}$}             & 20, 128, 64, 128, 4\\ 
\multicolumn{1}{c|}{$G_{1}$}             & 20, 128, 64, 128, 128, 64, 4\\ 
\multicolumn{1}{c|}{$G_{0}$}             & 20, 128, 64, 128, 128, 64, 4\\ 
\hline \hline
\end{tabular}
\caption{Configuration of the reduced CNN system. Convolution kernels with the size of $3\times 3 \times 3$ are utilized.}
\label{tab:reduced-CNN}
\end{table}

\begin{figure}
  \centering
  \subfigure[]{\includegraphics[width=0.35\linewidth,trim={0.60cm 0.60cm 0.60cm 0.20cm},clip]{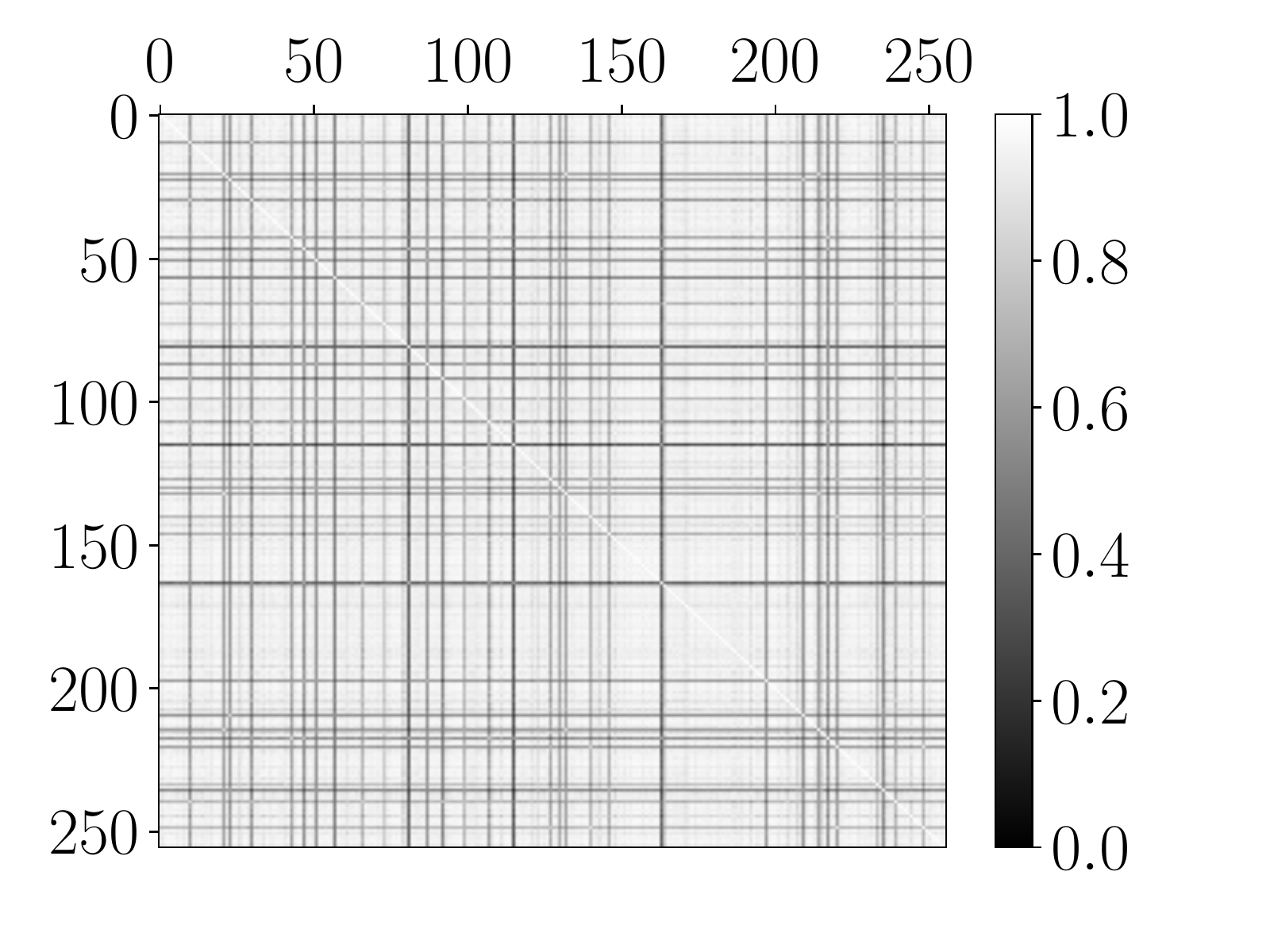}}
  \subfigure[]{\includegraphics[width=0.35\linewidth,trim={0.60cm 0.60cm 0.60cm 0.20cm},clip]{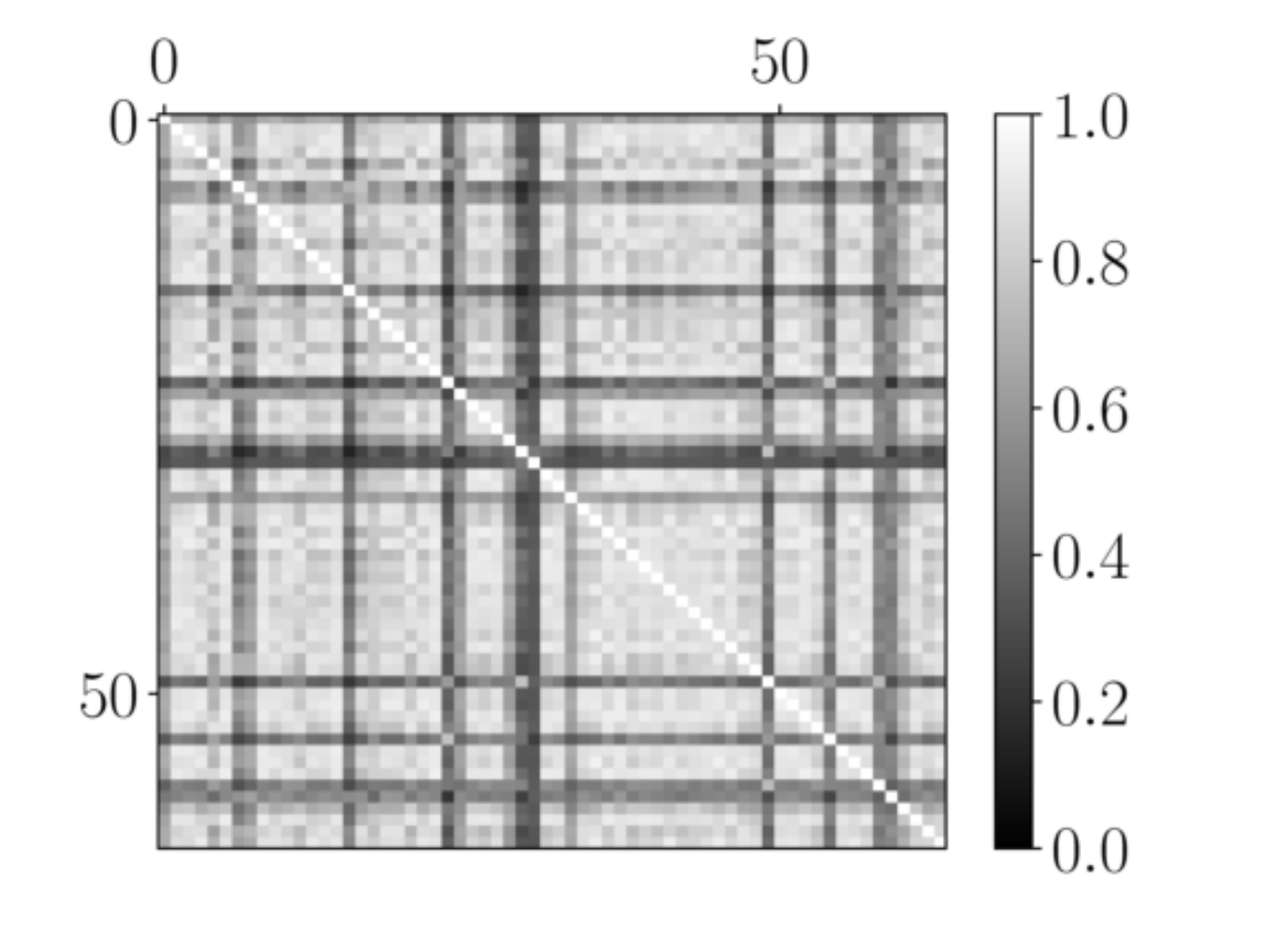}}

  \subfigure[]{\includegraphics[width=0.35\linewidth,trim={0.60cm 0.60cm 0.60cm 0.20cm},clip]{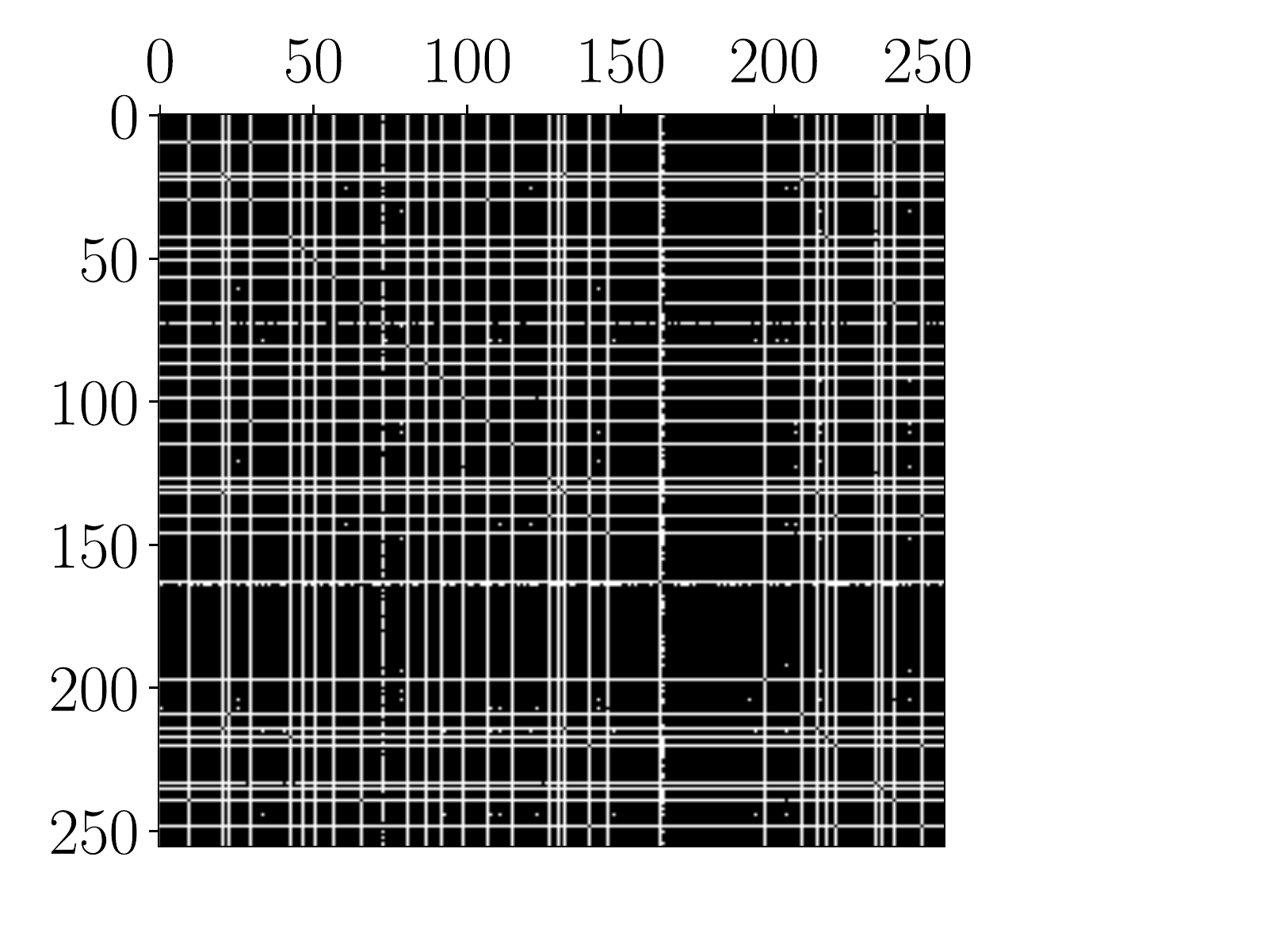}}
  \subfigure[]{\includegraphics[width=0.35\linewidth,trim={0.60cm 0.60cm 0.60cm 0.20cm},clip]{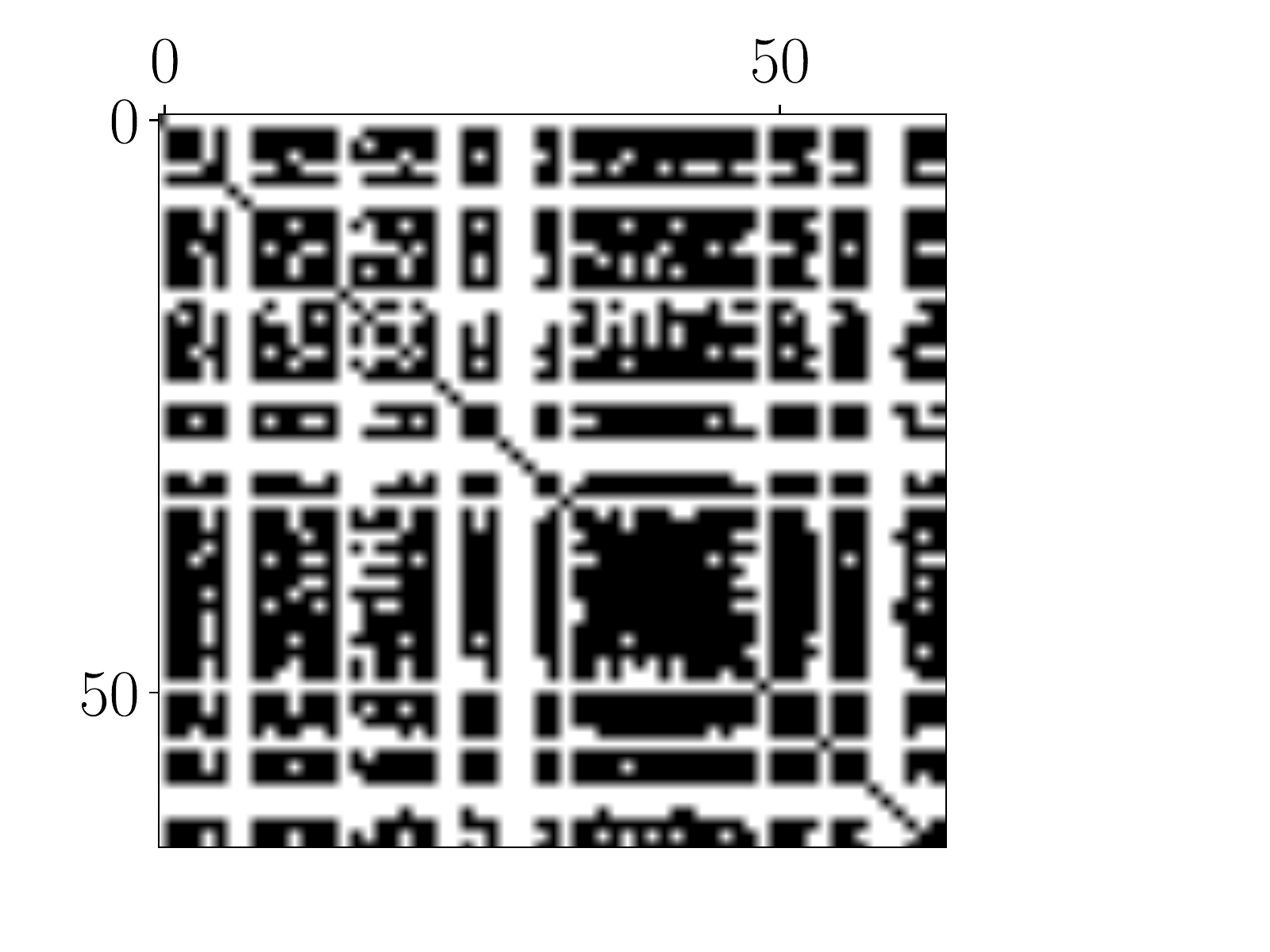}}
  \caption{Visualization of similarity matrices on the third layer of feature maps in $G_{0}$ of (a) the original CNN system and (b) the reduced CNN system. Elements with SSIM values over the threshold SSIM value of $0.80$ are masked (black colored). Masked elements in (c) the original CNN system and (d) the reduced CNN system are shown. The horizontal and vertical axes represent indices of feature maps.}
  \label{fig:sim}
\end{figure}

\begin{figure}
  \centering
  \subfigure[]{\includegraphics[width=0.30\linewidth,trim={0.0cm 0.0cm 0.0cm 0.0cm},clip]{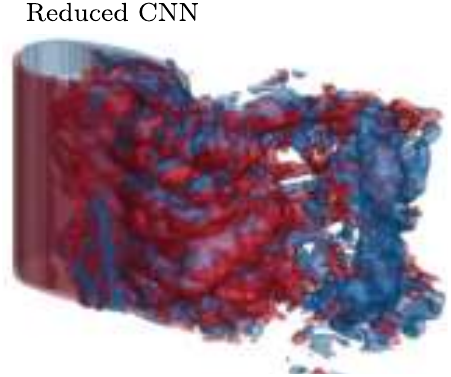}}

  \subfigure[]{\includegraphics[width=0.35\linewidth,trim={0.60cm 0.60cm 0.60cm 0.60cm},clip]{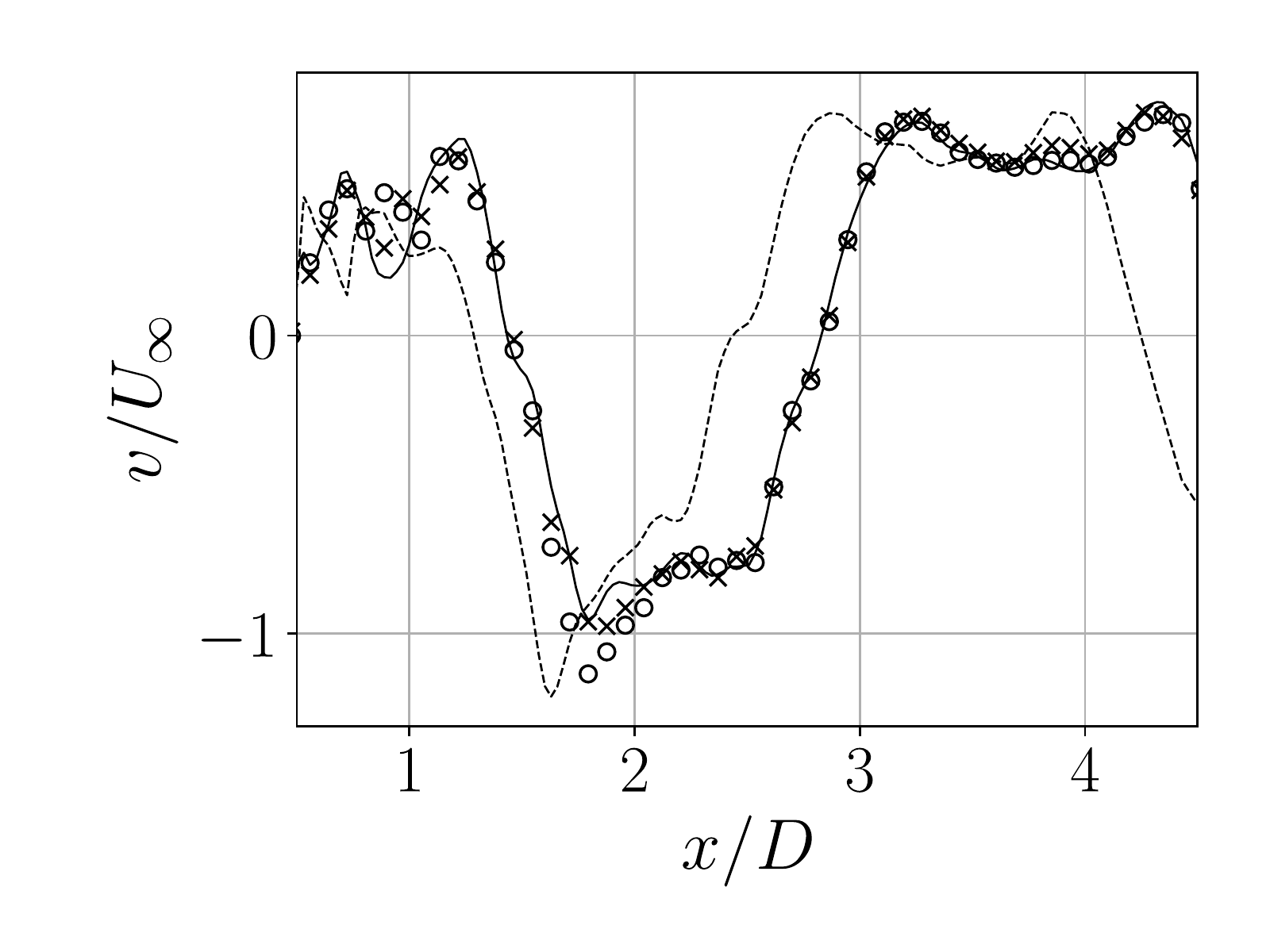}}
  \subfigure[]{\includegraphics[width=0.35\linewidth,trim={0.60cm 0.60cm 0.60cm 0.60cm},clip]{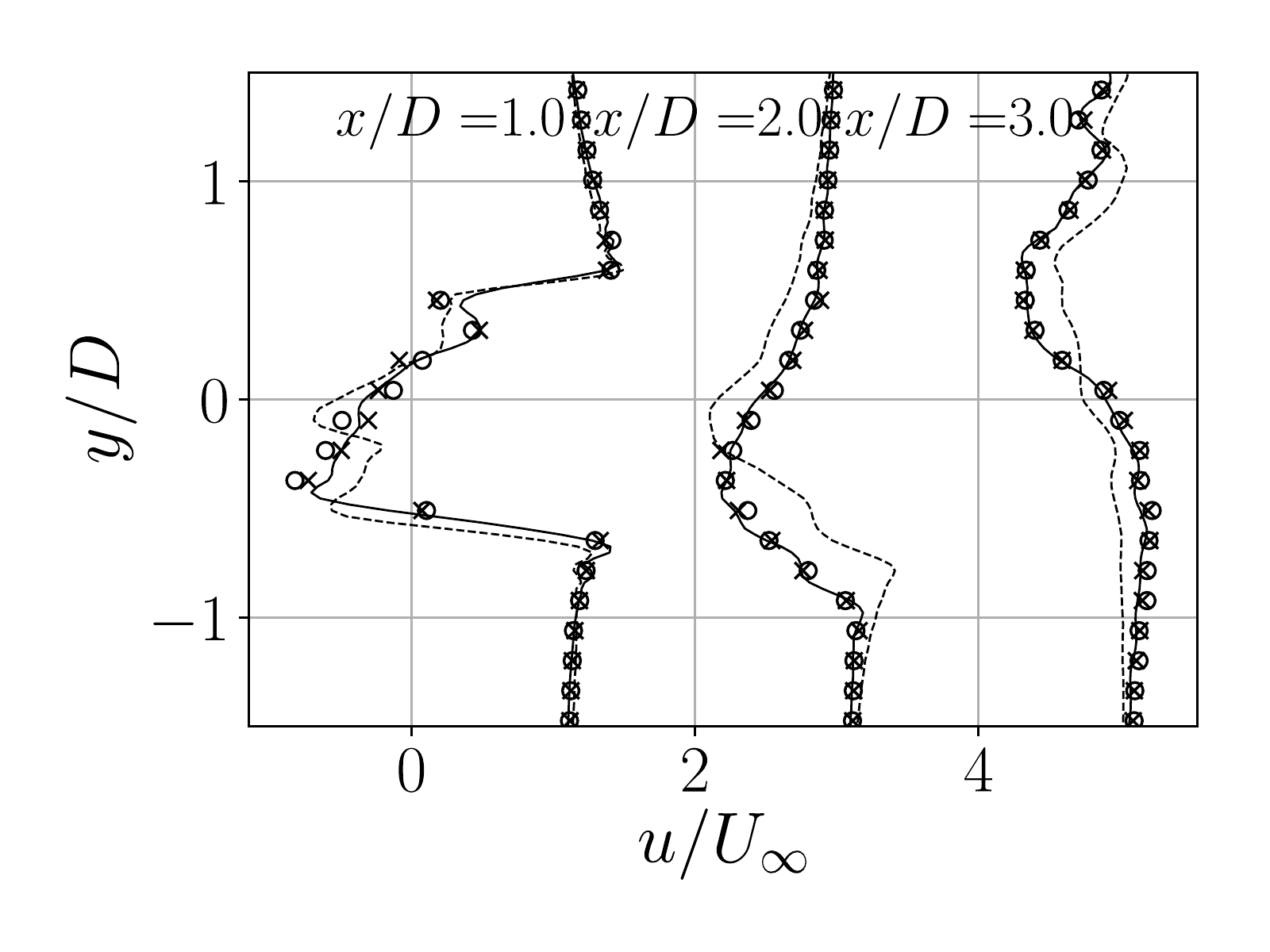}}
  \caption{(a) Isosurfaces of the instantaneous streamwise ($\omega_{x}D/U_{\infty}$) and spanwise ($\omega_{z}D/U_{\infty}$) vortices in the wake of a circular cylinder at $Re_{D}=3900$ and at $5\delta t$, calculated from the result predicted by the reduced CNN system. Red-colored isosurfaces, $\omega_{x}D/U_{\infty}=\omega_{z}D/U_{\infty} = 2.0$; blue-colored isosurfaces, $\omega_{x}D/U_{\infty}=\omega_{z}D/U_{\infty} = -2.0$. Profiles of (b) the instantaneous cross-stream velocity at $Re_{D}=3900$ along $x$ axis at $y/D=0$ and $z/D=1.5$, and (c) the instantaneous streamwise velocity along $y$ axis at $x/D=1.0,2.0,3.0$ and $z/D=1.5$. Dotted lines indicate profiles from input flow fields at $0\delta t$, solid lines indicate profiles from flow fields predicted by the reduced CNN system at $5 \delta t$. $\times$ and $\circ$ indicate profiles from flow fields predicted by the original CNN system and from the ground truth flow fields at $5 \delta t$, respectively.  Profiles of the streamwise velocity at $x/D=2.0$ and $3.0$ are shifted by $2.0$ and $4.0$ in the horizontal axis, respectively.}
  \label{fig:reduced-CNN}
\end{figure}

Similarity matrices calculated from the original CNN system and the reduced CNN system on the third layer of feature maps in $G_{0}$ are visualized in Figs.~\ref{fig:sim}(a)~and~(b).
Information of elements with SSIM values over the threshold SSIM value is provided in Figs.~\ref{fig:sim}(c)~and~(d) by masking elements in the similarity matrices.
It is found that the reduced CNN system contains smaller SSIM values compared to the original CNN system.
Especially, a decrement of proportion of masked elements, of which proportion corresponds to the $P(T)$ value in the calculated layer, is observed in the reduced CNN system.
Proportions of elements with SSIM values larger than the threshold SSIM value in all similarity matrices in the original CNN system and the reduced CNN system are compared by calculating the total proportion $P^{total}(T)$ as follows:
\begin{equation}
P^{total}(T) = \frac{count[(s,l,m,n) | S^{s,l,m,n} > T]}{\sum_{s,l,m,n}},
\label{eqn:Pl}
\end{equation}
where $S^{s,l,m,n}$ is the SSIM value between feature maps of $F^{m}$ and $F^{n}$ on the $l$th layer in the generative CNN of $G_{s}$, and $\sum_{s,l,m,n}$ is the sum of numbers of elements in all similarity matrices in the original CNN system or the reduced CNN system.
Values of $P^{total}(T)$ in the original CNN system and the reduced CNN system are calculated as $0.78$ and $0.49$, respectively. Therefore, the proportion of feature maps with high similarities is significantly reduced due to the use of the reduced CNN system, of which reduction indicates the removal of redundant feature maps.
As shown in Fig.~\ref{fig:reduced-CNN}, the reduced CNN system is found to be capable of generating flow structures with small length scales for flow at $Re_{D}=3900$ (Fig.~\ref{fig:reduced-CNN}(a)). Moreover, wake profiles predicted from the reduced CNN system are nearly identical to the results predicted by the original CNN system (Figs.~\ref{fig:reduced-CNN}(b)~and(c)).

\section{Conclusion}{\label{sec:conclusion}}
The predictive mechanisms of a CNN system which consists of multiple multi-resolution CNNs for predicting three-dimensional vortex dynamics in the wake of a circular cylinder have been investigated. 
The present CNN system is found to predict flow in different flow regimes and at different Reynolds numbers from flow with which the system was trained by using wavenumber information in input flow fields are transported through convolution layers. A convolution layer integrates wavenumber information extracted from convolution kernels and transports the integrated information to the next layer. By integrating and transporting the wavenumber information, the CNN extracts representations of flow dynamics in various length scales.

Feature maps which are outputs of composite functions of convolution operations, contain extracted representations of flow dynamics. Relative sizes of information in feature maps are found to be nearly unchanged even at different Reynolds numbers. Consequently, flow structures in feature maps calculated with different Reynolds numbers are similar regarding types of flow structures such as braid shear layers and shedding vortices. Smaller scale flow structures are additionally observed in feature maps calculated with flow at higher Reynolds numbers. Feature maps on a deeper layer show more clear types of flow structures compared to those on a shallow layer, because information from input flow variables and histories is integrated as the convolution layer goes deeper.

Contributions of input information to feature maps are evaluated to investigate the transportation characteristics of input flow variables and histories through the CNN system.
As the convolution layer goes deeper, nearly equal contributions from all input flow variables are found, which indicates that information from all flow variables is utilized to predict flow dynamics.
However, unequal contributions from input flow histories are observed in a deep layer of feature maps.
Contributions of information from the most recent flow history tend to be the largest, while contributions from other flow histories decrease as time distances to the prediction increase.
Therefore, the CNN system is considered to learn temporal correlations in unsteady flow dynamics as the temporal correlations generally decrease as time distances between flow fields increase.

A systematic approach to detect unnecessary feature maps in the CNN system is explored. Firstly, it is found that a feature map with a small relative size of information can contain clear flow structures affected by all flow variables and histories, and therefore, the number of feature maps cannot be reduced based on the size of information in feature maps. However, it is found that redundant feature maps, which contain similar flow structures, exist in layers of feature maps. The number of feature maps are found to be reduced by 85\%, for the present CNN system, based on evaluation of similarity matrices in layers of feature maps. 
The reduced CNN system shows indistinguishable prediction for the wake flow compared to that obtained by the original (before reduction of redundant feature maps) CNN system. The present finding is expected to be useful for deepening our understanding of the predictive mechanisms of CNN-based networks for learning fluid flow and for developing optimized networks with significantly reduced dependency on trial-and-error type efforts.

\section{Acknowledgements}\label{sec:Acknowledgements}
This work was supported by the Samsung Research Funding Center of Samsung Electronics under Project Number SRFC-TB1703-01 and National Research Foundation of Korea (NRF) under Grant Number NRF-2017R1E1A1A03070514.

%


\end{document}